\documentclass[11pt,a4paper]{iopart}
\usepackage{iopams}
\usepackage{graphicx,epsfig}% Include figure files
\usepackage{dcolumn}% Align table columns on decimal point
\usepackage{bm}% bold math
\newcommand{\be}{\begin{equation}}
\newcommand{\ee}{\end{equation}}
\newcommand{\bq}{\begin{eqnarray}}
\newcommand{\eq}{\end{eqnarray}}
\newcommand{\bc}{\begin{center}}
\newcommand{\ec}{\end{center}}

\begin{document}

%\begin{flushright}
%DAMTP-XXXX-YY\\
%\end{flushright}

\title[Constraints on the CMB temperature-redshift dependence]{Constraints on the CMB temperature-redshift dependence from SZ and distance measurements}

\author{A. Avgoustidis$^{1,2}$, G. Luzzi$^{3}$, C.J.A.P. Martins$^{4}$, A.M.R.V.L. Monteiro$^{4,5}$}
{\it 
$^1$School of Physics and Astronomy, University of Nottingham, University Park, Nottingham NG7 2RD, England\\
$^2$Centre for Theoretical Cosmology, DAMTP, CMS, Wilberforce Road, Cambridge CB3 0WA, England -- A.Avgoustidis@damtp.cam.ac.uk\\
$^3$Laboratoire de l'Acc\'el\'erateur Lin\'eaire, Centre Scientifique d'Orsay, B\^atiment 200 - BP 34, 91898 Orsay cedex - France -- gluzzi@lal.in2p3.fr\\
$^4$Centro de Astrof\'{\i}sica, Universidade do Porto, Rua das Estrelas, 4150-762 Porto, Portugal -- Carlos.Martins@astro.up.pt\\
$^5$Faculdade de Ci\^encias, Universidade do Porto, Rua do Campo Alegre 687, 4169-007 Porto, Portugal -- up090322024@alunos.fc.up.pt}

%\date{\today}

\begin{abstract}
The relation between redshift and the CMB temperature, $T_{\rm CMB}(z)=T_0(1+z)$ is a key prediction of standard cosmology, but is violated in many non-standard models. Constraining possible deviations to this law is an effective way to test the $\Lambda$CDM paradigm and search for hints of new physics. We present state-of-the-art constraints, using both direct and indirect measurements. In particular, we point out that in models where photons can be created or destroyed, not only does the temperature-redshift relation change, but so does the distance duality relation, and these departures from the standard behaviour are related, providing us with an opportunity to improve constraints. We show that current datasets limit possible deviations of the form $T_{\rm CMB}(z)=T_0(1+z)^{1-\beta}$ to be $\beta=0.004\pm0.016$ up to a redshift $z\sim 3$. We also discuss how, with the next generation of space and ground-based experiments, these constraints can be improved by more than one order of magnitude.
\end{abstract}

%\pacs{}
%\keywords{}
%\preprint{}

%------------------------------------------------------------------------------------------------------------------------

\section{\label{intro} Introduction}

Cosmology and particle physics are presently experiencing a truly exciting period. On the one hand, both have remarkably successful standard models, which are in agreement with a plethora of experimental and observational data. On the other hand, there are also strong hints that neither of these models is complete, and that new physics may be there, within the reach of the next generation of probes.

There are three compelling and firmly established observational facts that the standard model of particle physics fails to account for: neutrino masses, the existence of dark matter, and the size of the baryon asymmetry of the Universe. For each of these, the model makes very specific statements, failing however to reproduce the experimental evidence. It is precisely our confidence in the model and our ability to calculate its consequences that lead us to the conclusion that it is incomplete, and new phenomena must be anticipated. This is, of course, the reason for the LHC project.

Similarly, the last decade saw the emergence of the so-called concordance model of cosmology. This can reproduce all the available observations with only a small number of parameters, but also requires that about $96\%$ of the content of the universe is in a form that has never been seen in the laboratory (and is only known indirectly from its gravitational properties). It is thought that dark matter is a subdominant part of this, while the dominant one is an even more mysterious component usually called dark energy.

In this context, it is important to identify laboratory or astrophysical probes that can give us more information about the nature and properties of this still unknown physics. In this work we will discuss one such probe---the temperature-redshift relation---, and lay the foundations for exploring its cosmological implications.

One of the most precise measurements in cosmology is the intensity spectrum of the cosmic microwave background radiation: the COBE-FIRAS experiment revealed a very precise black-body spectrum \cite{Mather}. However, this measurement tells us nothing about the behaviour of the Cosmic Microwave Background (CMB) at non-zero redshift. If the expansion of the Universe is adiabatic and the CMB spectrum was a black-body at the time it originated, this shape will be preserved with its temperature evolving as $T(z)=T_0(1+z)$. This is a robust prediction of standard cosmology, but it is violated in many non-standard models, including string theory motivated scenarios where photons mix with other particles such as axions (see~\cite{JaeckRing} for a recent review), and those where dimensionless couplings like the fine-structure constant vary \cite{RoySoc}.

A few measurements of $T(z)$ already exist, but the currently large uncertainties do not allow for strong constraints on the underlying models to be set. However, with future datasets this will become a competitive probe. It is therefore timely to discuss what these measurements can tell us about the underlying cosmological paradigms.

At low redshifts, say $z<1$, the $T(z)$ relation can be measured via the Sunyaev-Zeldovich (SZ) effect towards galaxy clusters. This method was applied to ground-based CMB observations \cite{Battistelli,Luzzi}, which demonstrated its potential. With a new generation of ground experiments becoming operational and a forthcoming all-sky survey of SZ clusters to be carried out by Planck \cite{Horellou}, the potential of this method will come to fruition. At higher redshifts, $z>1$, $T(z)$ can be evaluated from the analysis of quasar absorption line spectra which show atomic and/or ionic fine structure levels excited by the photon absorption of the CMB radiation \cite{Srianand}. (The CMB is an important source of excitation for species with transitions in the sub-millimeter range.) Although the suggestion is more than four decades old, measurements (as opposed to upper bounds) were only obtained in the last decade, and the best ones so far still have errors at the ten percent level \cite{Noterdaeme}.

Here we will study these issues in detail, but we will also place them in a wider context. For example, in models where photons can be created or destroyed, not only does the temperature-redshift relation vary, but so does the distance duality relation (also known as the Etherington relation \cite{Etherington1}), and these two different departures from the standard behaviour are quantitatively related. One issue that has been overlooked so far is that in such models, where photon number is not conserved, this relation between $T(z)$ and distance duality provides us with an opportunity to improve constraints. By combining data from different observations one not only reduces the statistical uncertainties on underlying phenomenological parameters but, given the different nature of both observational datasets, one also has a much better control over possible systematics.

We therefore discuss in detail the origin of the above relation, as it can be a unique consistency test for the standard paradigm and, at the same time, a valuable tool for probing new physics beyond the standard model. We also study further imprints of these models in the CMB, and present forecasts for improvements that Planck, as well as planned Baryon Acoustic Oscillations (BAO) missions and spectrographs planned for the VLT and the E-ELT, will soon make possible. Last but not least, we derive the strongest constraints to date on deviations of these relations from their standard behaviour, and quantify the improvements to be expected from the aforementioned forthcoming experiments.

\section{\label{theory} Theoretical motivation}

There are several examples of non-standard, but theoretically well-motivated, physical processes that could affect the cosmological temperature-redshift relation.  Constraining deviations from the standard 
law $T(z)=T_0(1+z)$ therefore provides a invaluable tool for probing physical theories. Examples of scenarios that could be constrained include decaying vacuum cosmologies/photon injection mechanisms, couplings between photons and axion-like particles, modified gravity scenarios, and so on. In this section, we discuss two examples of how these models affect the $T(z)$ relation. 

\subsection{Adiabatic Photon Injection\label{PhotonInjection}}

Naively, the most obvious way to violate the $T(z)=T_0(1+z)$ relation is when there is energy injection into the CMB (say from a decaying scalar field) or, conversely, when photons are destroyed. The Planckian form of the spectrum is preserved if photon creation is adiabatic, that is, if the entropy per photon remains constant. Lima {\it et al.} \cite{Lima1,Lima2} have studied this in the context of decaying vacuum cosmologies; here we review some of their formalism.

Generically, such a process of particle creation can be described by saying that the energy density evolves as
\begin{equation}
{\dot\rho}+3H(\rho+p)=C
\end{equation}
while the particle number density obeys
\begin{equation}
{\dot n}+ 3H n=\Psi\, ,
\end{equation}
where $C$, $\Psi$ are real functions of time.
Deviations from particle number conservation can then be quantified by the phenomenological parameter
\begin{equation}
0\le\beta\equiv\frac{\Psi}{3H n}\le1\,, \label{defbeta}
\end{equation}
which can be time-dependent.

In general the temperature evolves as
\begin{equation}
\frac{\dot T}{T}=\left(\frac{\partial p}{\partial \rho}\right)_{\!\!n}\frac{\dot n}{n}-\frac{\Psi}{nT\left(\partial\rho/\partial T\right)_n}\left[\rho+p-\frac{nC}{\Psi}\right]\, ,
\end{equation}
but note that the second term is zero if $C$ and $\Psi$ are related by
\begin{equation}
C=\frac{\rho+p}{n}\Psi \,,
\end{equation}
or, in terms of the parameter defined in (\ref{defbeta}), $C=3H(\rho+p)\beta$.  This corresponds to adiabaticity: the specific entropy per particle 
of the created particles remains constant, so new particles are actually created in equilibrium with 
already existing ones.  In this case the temperature evolution equation reduces to
\begin{equation}
\frac{\dot T}{T}=\left(\frac{\partial p}{\partial \rho}\right)_{\!\!n}\frac{\dot n}{n} \label{Tdotsimple}\,,
\end{equation}
as in the standard cosmological model. Recall that, in the standard model, both the equilibrium relations and the Planckian form of the spectrum for photons are preserved in the course of expansion. The former can be seen directly from equation (\ref{Tdotsimple}), while the latter is the result of the kinematical condition for FRW geometry, $\nu\propto a^{-1}$, and of photon number conservation, which together imply $T\propto a^{-1}$ (away from mass thresholds). With adiabatic photon creation, equation (\ref{Tdotsimple}) remains valid -- as we just saw -- and thus the equilibrium relations are preserved. 

Indeed, if one considers the generic equation of state $p=(\gamma-1)\rho$ one can integrate the above equations to get
\begin{equation}
T\propto n^{\gamma-1}\,,\quad \rho \propto n^\gamma \,.
\end{equation}
However, since the entropy per photon is conserved, the temperature now obeys
\begin{equation}
\frac{Ta}{N^{1/3}}={\rm const} \,,
\end{equation}
where $N(z)$ is the comoving photon number, which changes in time as new photons are being injected in equilibrium.  It then follows that the dimensionless frequency relevant for the SZ effect (to be considered in \S\ref{SZ}) evolves as
\begin{equation}
x=x_0(1+z)\frac{T_0}{T_{\rm CMB}} \,,
\end{equation}
with
\begin{equation}
T_{\rm CMB}=T_0(1+z)\left[\frac{N(z)}{N_0}\right]^{1/3} \,.
\end{equation}
Thus, it is now $h\nu(N/N_0)^{1/3}/kT$ in the exponential of the photon distribution 
function that stays constant during cosmological evolution, and so a generalised Planck-type 
spectrum~\cite{Lima1} is preserved, which is why the equilibrium relations are still recovered, e.g. $\rho\propto T^4$
for a radiation fluid (where the proportionality factor is the radiation density constant). 

Although one cannot distinguish at present the usual Planckian spectrum from such a generalised one, this is not necessarily true at higher redshifts. For example, the wavelength $\lambda_m$ of the peak of the distribution is now
\begin{equation}
\lambda_m\,T=0.289\left[\frac{N(z)}{N_0}\right]^{1/3}\,,
\end{equation}
which is a generalisation of Wien's law (and naturally reduces to it in the standard case). This could in principle be observationally tested.

We can illustrate these points with two simple examples. Assuming a radiation fluid, $p=\rho/3$ we have
\begin{equation}
\frac{\dot T}{T}=-H+\frac{\Psi}{3n}=-H+\frac{C}{4\rho}
\end{equation}
and for a constant $\beta$ we find
\begin{equation}
{\dot\rho}+4(1-\beta)H\rho=0
\end{equation}
\begin{equation}\label{ndotbeta}
{\dot n}+3(1-\beta)H n=0 
\end{equation}
\begin{equation}\label{Tofz}
T(z)=T_0(1+z)^{1-\beta} \,;
\end{equation}
the dimensionless frequency is therefore
\begin{equation}
\frac{x}{x_0}=(1+z)^\beta\,.
\end{equation}
For a more general equation of state $p=(\gamma-1)\rho$ we have
\begin{equation}
T\propto a^{-3(\gamma-1)(1-\beta)}\,.
\end{equation}
Thus, for any given redshift the temperature of the expanding universe is slightly lower than in the standard case. This is the phenomenological parameterisation that has been used by almost all previous authors. Note that for many realistic models this is not an accurate description (as $\beta$ is in general time-dependent), but it is adequate at sufficiently low redshifts. Since, currently, cluster data probe down to redshifts less than unity with sensitivities in temperature of order a few percent or worse (see \S\ref{SZ}), we will adopt this parameterisation for the purposes of the present paper (but will revisit the issue in a follow-up publication). As cosmological data improve and extend to higher redshifts (see \S\ref{forecasts}), time variation of the parameter $\beta$ will also be constrained.

\subsection{Photon dimming/absorption \& Axion-Photon Couplings\label{PhotonDimming}}

The effect of photon dimming/absorption can arise in a wide range of astrophysical 
and high-energy physics scenarios, ranging from photon absorption by grey dust 
to photon conversion into a different particle species, like for example photon-to-axion 
conversion in the presence of a magnetic field.

Grey dust has been invoked as an 
alternative explanation of the observed dimming of Type Ia Supernovae \cite{Aguirre}, 
but it is now understood that it cannot fit high redshift data (like for example the Union
sample \cite{Union}) by itself, i.e. without some contribution from a cosmological 
constant-like fluid \cite{AVJ}.  Axion-Like-Particles (ALPs), on the other hand, arise in 
a wide range of well-motivated high-energy physics scenarios, including string theoretic
models where ALPs appear as zero modes of various antisymmetric form fields \cite{SvrWitt}.  
Like dust, they can lead to dimming of Type Ia Supernovae~\cite{Csaki}, but this cosmological 
scenario is also strongly constrained~\cite{OstmMorts,MirRafSerp,ABRVJ}.  For a recent 
review of ALPs and related laboratory, astrophysical and cosmological constraints, 
see \cite{JaeckRing}.    

Irrespectively of its microphysical origin, photon dimming violates photon number 
conservation, and so it can be described macroscopically in a very similar way 
to photon injection, by simply allowing the parameter $\beta$ in equation 
(\ref{defbeta}) to be negative.  In particular, the balance (\ref{ndotbeta}) and temperature 
evolution (\ref{Tdotsimple}) equations are valid with constant negative $\beta$ (note
that, as before, constant $\beta$ is only an approximation) and the temperature-redshift
relation is again:
\be
T(z)=T_0 (1+z)^{1-\beta} \,.
\ee       
 
Crucially for this work, violations of photon number conservation also give 
rise to deviations from the standard relation~\cite{Etherington1} between 
luminosity and angular diameter distance, which can be independently 
constrained \cite{bassettkunz1,bassettkunz2,More,AVJ,ABRVJ}.  With current 
data, it is sufficient to parameterise the luminosity-angular diameter distance 
relation as:      
\be\label{dLdA_eps}
d_L(z) = d_A(z) (1+z)^{2+\epsilon}\,,
\ee    
where $\epsilon$ is constant. The value $\epsilon=0$ corresponds to the 
standard $d_L-d_A$ law.  Since $\epsilon$ and $\beta$ are related 
by the underlying physical model but can be constrained independently 
through measurements of very different systematics, this provides a
very promising tool for carrying out a consistency test of the standard
cosmological scenario and constraining physics beyond the standard model. 
However, such complementary measurements often cover a very different 
wavelength range so one must be careful when comparing direct and 
indirect constraints.  For example, if one uses Supernova (SN) data to 
constrain the parameter $\epsilon$ at optical wavelengths within a model 
of chromatic axion-photon mixing, one should take into account the 
wavelength dependence of the model when translating this bound 
into a constraint in the CMB temperature-redshift relation.  In general,
one may expect photon-dimming to be stronger at high photon energies, 
so indirect bounds on $T(z)$ coming from SN brightness measurements  
can be assumed to be conservative.  In \S\ref{distance} we will present 
the constraints obtained from such an analysis with current data, and 
in \S\ref{forecasts} we examine the prospects of improving these 
constraints by future SN and BAO data.  

In passing, we note that another way to see that the temperature-redshift relation 
becomes modified in models violating photon number conservation 
is by considering the more obvious effect of the distance-duality 
violation in these models.  The source of this distance-duality violation 
is the difference between the true and observed (dimmed) photon flux, 
which tricks one to infer a larger luminosity distance.  As flux scales 
inversely with the luminosity distance squared, this change in the inferred 
distance would also affect the CMB flux-redshift relation, and so the CMB 
temperature as a function of redshift.

Before moving to the study of the various constraints imposed on 
$\beta$ and $\epsilon$, let us explore their relation in more detail.  Consider the CMB 
spectrum in the presence of a dimming agent, like for example photon 
absorption due to a dust field or conversion into ALPs.  Imagine that the 
spectrum is Planckian at the epoch $T$ so that the the number of photons 
per unit volume per frequency interval is:
\be\label{photon_dens}
n(\nu)d\nu=\frac{8\pi}{c^3} \frac{\nu^2 d\nu}{e^{h \nu/kT}-1}\,.
\ee 
If photon number was conserved, then at a later epoch $T'$ the 
number density per frequency interval, $n'(\nu')d\nu'$, would be related 
to (\ref{photon_dens}) through the volume rescaling $({a/a'})^3$;  
this can be absorbed as frequency dependence, 
$\nu'=(a/a')\nu$, yielding the standard result of a Planckian spectrum 
$n(\nu')d\nu'$ of temperature $T'=(a/a')T$.  With dimming, however, photons 
can be lost in flight and the spectrum will generally be distorted.   Introducing  
a photon survival fraction $f_{\rm surv}\lesssim 1$ between the epochs $T$ 
and $T'$ we now have:
\[
n'(\nu')d\nu' = f_{\rm surv}\, n(\nu)d\nu \left({a/a'}\right)^3 =
\frac{8\pi}{c^3} \frac{f_{\rm surv}\, \nu'^2 d\nu'}{e^{h \nu'/kT'}-1}\lesssim 
n(\nu')d\nu' \,,
\]   
where, as before, $\nu'=\nu(a/a')$ and $T'=(a/a')T$.  Thus the spectrum 
is distorted.  However, if the photon survival fraction does not depend on 
frequency (of course, it can---and generally will---depend on redshift), 
then one can change variables to $\nu''=(f_{\rm surv})^{1/3}\,\nu'$\,, yielding:  
\[
n'(\nu')d\nu' = n(\nu'') d\nu''\,,
\] 
that is, a Planckian spectrum of temperature $T''=(f_{\rm surv})^{1/3}\,T'
=(a/a') (f_{\rm surv})^{1/3}\,T$.  This is the `generalised' Planck distribution 
of reference \cite{Lima1}, discussed in the previous section.  In fact, remembering
that $T'$ corresponds to an epoch later than $T$, we have:
\be 
T(z)=(f_{\rm surv})^{-1/3}(1+z)T_0 \,.
\ee

The relation between dimming and temperature distortion then becomes
clear.  In the parameterisation (\ref{Tofz}), we have 
$(f_{\rm surv})^{1/3}=(1+z)^\beta$, with $\beta$ negative.  On the other 
hand, the photon survival probability $f_{\rm surv}$ enters linearly the 
luminosity, so the luminosity distance scales as $(f_{\rm surv})^{-1/2}$.  
Therefore, from equation (\ref{dLdA_eps}) it follows that the photon survival 
probability corresponds, in the $\epsilon$-parameterisation, to 
$f_{\rm surv}=(1+z)^{-2\epsilon}$.  Thus, the relation between the parameter 
$\beta$ of equation (\ref{Tofz}) and $\epsilon$ of references \cite{AVJ,ABRVJ} 
is simply:
\be\label{beta_eps}
\beta=-\frac{2}{3}\epsilon \,.
\ee          
      
Note that in the above discussion we have assumed that the photon survival 
probability is independent of wavelength, which guaranteed a simple thermal 
spectrum.  Deviations of the CMB from a thermal spectrum have been constrained 
down to the $10^{-4}$ level by the COBE-FIRAS measurements, both in the case 
of a Comptonised spectrum and of a Bose-Einstein spectrum with non-trivial 
chemical potential \cite{Mather,FIRAS}.  This dataset has also been used recently 
to constrain frequency-dependent dimming in the context of photon-axion couplings
\cite{MirRafSerp}.  Finally, we have assumed a simple
redshift dependence, parameterised by equation (\ref{dLdA_eps}) with constant 
$\epsilon$.   Again, like in the case of constant $\beta$ (section \ref{PhotonInjection}),
this is justified given current error bars in distance determination and the limited redshift 
range currently covered, $0\lesssim z\lesssim 2$ \cite{AVJ}.  Forecasts for future data will be 
discussed in \S\ref{forecasts}. 

\section{\label{SPC} Constraints from quasar absorption line spectra}

The most accurate value of the local CMB temperature measured by the COBE experiment is \cite{Mather}
\be
T_{\rm CMB} = 2.725\pm0.002\, K\,,\quad z=0\,.
\ee
Additional local but extrasolar values of the background radiation temperature can be estimated from observations of interstellar molecular clouds \cite{Roth} and of the Magellanic Clouds \cite{Welty}. These extrasolar and extragalactic values are in good agreement with the COBE estimation. Cosmological models predicting non-Planckian spectra of the background radiation at $z  = 0$ can be ruled out if they deviate by more than about $1\%$ from the blackbody spectrum. However, these local observations do not allow one to distinguish between the standard model and cosmological models with a blackbody spectrum but different $T(z)$ dependences.  

At low redshifts ($0 < z <1$), the functional scaling of the CMB temperature can be estimated from measurements of the Sunyaev-Zeldovich effect in the direction to clusters of galaxies, as will be discussed in the next section. At higher redshifts ($z>1$), the CMB temperature can be evaluated from the analysis of quasar absorption line spectra which show atomic and/or ionic fine structure levels excited by the photo-absorption of the CMB radiation. This is an important source of excitation for those species with transitions in the sub-millimetre range. This is the case for atomic species whose ground state splits into several fine-structure levels (and also for molecules that can be excited in their rotational levels, see below). If the relative level populations are thermalised by the CMB, then the excitation temperature gives the temperature of the black-body radiation.

It has long been proposed to measure the relative populations of such atomic levels in quasar absorption lines to derive $T_{\rm CMB}$ at high redshift \cite{Bahcall}. The most suitable are the fine structure levels of the ground states of CI ($3P0,1,2$) and CII ($2P1/2,3,2$) showing an energy separation between 24 K and 91 K. The CI lines were used to obtain several upper limits, until advances in instrumentation and analysis techniques allowed actual measurements, starting with Srianand {\it et al.} \cite{Srianand} who obtained
\be
T_e = 10\pm4\, K\,,\quad z = 2.338\,.
\ee
To distinguish the contribution to the relative population of the fine-structure levels of the ground state of CI from competing excitation processes the independent analysis of the molecular hydrogen UV absorption lines is important; $H_2$ transitions from different low rotational levels may be used to infer the UV radiation field and the gas density in the CI-$H_2$ absorbers. These techniques have so far allowed measurements to be made beyond $z=3$; the earliest currently available measurement is that of Molaro {\it et al.} \cite{Molaro}
\be
T_e = 12.1^{+1.7}_{-3.2}\, K\,,\quad z = 3.025\,.
\ee

Molecular rotational transitions may also be used to constrain the cosmological temperature-redshift law. So far, absorption and emission lines of $CO$, $OH$, $CS$, $HCN$, $HCO^+$, $H_2O$, $NH_3$, and other molecules have been observed in distant galaxies and quasars up to $z=6.42$ (see \cite{Omont} for a review). The absorption line observations are most suitable to the radiation temperature estimates since, as a rule, molecular absorption arises in the gas components with the lowest kinetic temperatures.  Srianand {\it et al.} \cite{SrianandCO} reported the first detection of CO in a high-redshift damped Lyman-$\alpha$ system, while also detecting $H_2$ and $HD$ molecules. The CO rotational excitation temperatures are higher than those measured in our Galactic ISM for similar kinetic temperature and density. Using the $CI$ fine-structure absorption lines, they show that this is a consequence of the excitation being dominated by radiative pumping by the cosmic microwave background radiation, and from the CO excitation temperatures they derive
\be
T_e = 9.15\pm0.72\, K\,,\quad z = 2.418\,.
\ee

Finally, Noterdaeme {\it et al.} \cite{Noterdaeme} have recently reported on a sample of five CO absorption systems where the CMB temperature has been measured. We refer the reader to this work for further details as well as for some more on the history of these measurements and an up-to-date list of all the available ones. They also used their sample, in combination with measurements from the SZ effect, to place constraints on the phenomenological parameter $\beta$---we will describe this in the next section.

\section{\label{SZ} Constraints from the SZ effect towards clusters}

The possibility of determining $T_{\rm CMB}(z)$ from measurements of the
Sunyaev-Zeldovich effect has been suggested long ago
\cite{fabbri1978, reph1980}. The effect -- Compton scattering of
the CMB by hot intracluster (IC) gas -- is a small change of the
CMB spectral intensity, $\Delta I$, which depends on the
integrated IC gas pressure along the line of sight to the
cluster. The steep frequency dependence of the change in the CMB
spectral intensity, $\Delta I$, due to the SZ effect allows the
CMB temperature to be estimated at the redshift of the cluster.

The differential SZ signal
may be written at the redshift of the cluster, including
relativistic corrections as:
\begin{equation}
\Delta I(z)=\frac{2(kT_{\rm
CMB}(z))^3}{(hc)^2}\frac{x^4e^x}{(e^x-1)^2}\tau\left[\theta
f_1(x)-v_z/c+R(x,\theta,v_z/c)\right] ,
\end{equation}
where $\Delta I(z)$ is the brightness change between the centre of
the cluster and blank sky, as measured at redshift $z$, $\tau$ is the optical depth, $T_{\rm
CMB}(z)$ is the CMB temperature at redshift $z$, $x=\frac{h\nu(z)}{k_BT_{\rm  CMB}(z)}$, 
$\theta=k_BT_e/m_ec^2$ with $T_e$ electron cluster temperature, $v_z$ the radial component 
of the peculiar velocity of the cluster, and the $R(x,\theta,v_z/c)$ function includes relativistic corrections. 

If we assume that $T_{\rm CMB}$ scales with $z$ as $T_{\rm CMB}(z)=T_{\rm
CMB}(0)(1+z)^{1-\beta}$, while the frequency scales as
$(1+z)$ as usual, then $\Delta I(z)$ scales like
$(1+z)^{3(1-\beta)}$ and we obtain, at the level of the solar system
\begin{equation}
\Delta I(0)=\frac{2(kT_{\rm CMB}(0))^3}{(hc)^2}\frac{x^{\prime
4}e^{x^{\prime}}}{(e^{x^{\prime}}-1)^2}\tau\left[\theta
f_1(x^{\prime})-v_z/c+R(x^{\prime},\theta,v_z/c)\right] ,
\end{equation}
where $x^{\prime}=\frac{h\nu(0)}{k_B T^{\ast}_{\rm CMB}}$ and
$T^{\ast}_{\rm CMB}=T_{\rm CMB}(0)(1+z)^{-\beta}$ will be slightly
different from the local temperature $T_{\rm CMB}(0)$ as measured
by COBE. In this way it is possible to measure the temperature of
the CMB at the redshift of the cluster, thus directly 
constraining scenarios like those discussed in the previous section.  

Let us consider the bound on $\beta$ coming from $T_{\rm CMB}(z)$ constraints, at redshifts 
in the range $z=0.023-0.546$, from multi-frequency measurements of the
SZ effect towards the 13 clusters of Ref \cite{Luzzi}.
By fitting the $T_{\rm CMB}(z)$ data points and relaxing the condition of a positive prior on $\beta$, that is allowing for photon dimming/absorption (corresponding to $\beta<0$) as well as photon creation ($\beta>0$), we get $\beta=0.065\pm0.080$.

By adding these measurements to higher-redshift ones coming from spectroscopic measurements (discussed in the previous section) involving atomic carbon and CO absorption lines along the line of sight of quasars, Ref. \cite{Noterdaeme} subsequently improved the constraints on $\beta$, obtaining\footnote{Repeating their fit to parameter $\beta$ we find an uncertainty of 0.028 rather than 0.027; this is likely due to round-off errors in the temperature measurements used as input. The difference is insignificant for the purposes of our subsequent analysis, but for the sake of consistency with the pipeline for comparing to future measurements we will use 0.028 in Figs.~\ref{fig:codex} and \ref{fig:allsigmas}.}
\begin{equation}\label{betadirect}
\beta=-0.007\pm0.027\,.
\end{equation}
Despite the higher precision of $T_{\rm CMB}(z)$ measurements from the SZ method with respect to that of the 
quasar absorption lines method, the present improvement on the constraints on the  parameter $\beta$ depends 
mainly on the higher lever arm due to the exploration of the distant universe, thanks to the observation of 
high redshift absorbers. As we will show in \S\ref{forecastsPlanck} forthcoming datasets will soon lead to tighter constraints.

\section{\label{distance} Constraints from distance measurements}

As briefly alluded to in \S\ref{theory}, an indirect way for constraining 
the temperature-redshift dependence is through the study of possible 
deviations from the duality relation~\cite{Etherington1} between luminosity 
and angular diameter distance:
\be\label{Etherington}
d_L(z)=(1+z)^2 d_A(z)\, .
\ee    
This equation holds for general metric theories of gravity, where 
photons travel along unique null geodesics, as long as photon number 
conservation and local Lorentz invariance are respected.  Since Lorentz 
violations are strongly constrained at low energies~\cite{Kostelecky:2008ts}, 
and in particular at optical wavelengths, the determination of $d_L$ from SN 
observations can be used~\cite{bassettkunz1,bassettkunz2,More,AVJ,ABRVJ}
together with other distance measurements in order to place direct bounds 
on photon number violation through Eq.~(\ref{Etherington}).  Physically, such a 
violation can arise from photon absorption, e.g. grey dust~\cite{Aguirre}, or from 
more exotic effects like photon conversion into axions~\cite{Csaki}.    
 
Systematic violations of (\ref{Etherington}) give rise to an apparent opacity 
effect in the observed luminosity distance.  Indeed, if photons were 
lost along the line of sight, the inferred luminosity distance would be related 
to the true one through a multiplicative factor:
\be\label{tau} 
d^2_{L,inf}=d^2_{L,true} e^{\tau(z)}\, .
\ee 
Note that the `opacity' $\tau(z)$ can be negative, allowing 
for apparent brightening of the source, as would be the case, for 
example, if exotic particles were also emitted from the source and 
later converted into photons \cite{Burrage:2007ew}. 

As discussed in \S\ref{theory}, photon number violation would also 
give rise to a corresponding distortion of the photon temperature-redshift 
relation, and this allows us to combine different observational probes to 
constrain such models. 

In references \cite{AVJ,ABRVJ} the authors used Type Ia SN data (specifically the Union 
2008 dataset~\cite{Union}) in combination with measurements of cosmic expansion 
$H(z)$ from differential ageing of luminous red galaxies \cite{JVST,SVJ05,SJVKS}, 
and obtained constraints on opacity up to $z\simeq 2$ through equations 
(\ref{Etherington}-\ref{tau}).  In the simplest case, one can adopt the 
parameterisation: 
\be\label{epsilon}
d_L(z) = d_A(z) (1+z)^{2+\epsilon} \, ,
\ee   
allowing violations of Eq.~(\ref{Etherington}) through a single parameter,
$\epsilon$.  For low redshifts, this parameter can be directly translated 
into an opacity function $\tau(z)$ as it simply corresponds to the first 
term in the Taylor expansion $\tau(z)=2\epsilon z + {\cal O}(\epsilon z^2)$.    
Different parameterisations were also considered in~\cite{ABRVJ}, corresponding
to specific theoretical models of opacity such as photons mixing with massless 
axion-like particles, chameleons and mini-charged particles.     

If an opacity source like photon-axion mixing affects SN observations, it 
should also have an impact on CMB photons.  From the above discussion 
(and that of \S\ref{PhotonDimming}) 
it is then clear that constraints like those in~\cite{AVJ,ABRVJ} can be used 
to place indirect bounds on possible deviations from the standard 
temperature-redshift law, as expressed, for example, in Eq.~(\ref{Tofz}).  
In particular, if opacity is achromatic, which can be the case, for example, 
for photon-axion mixing in sufficiently low intergalactic electron densities~\cite{Csaki2}, 
the parameterisation (\ref{epsilon}) corresponds simply to $\beta=-2\epsilon/3$, as
we saw in \S\ref{PhotonDimming}.  More generally, however, the effective opacity 
can be expected to be smaller at low energies, so that distance duality violation 
bounds yield conservative constraints for the allowed CMB temperature-redshift 
distortion. There are also other bounds on photon-axion mixing, coming from constraining 
spectral distortions of the CMB radiation~\cite{MirRafSerp}.  Note that for CMB 
photons to be converted into axions, a background magnetic field is required at the 
corresponding redshift.  Therefore, the bound on the parameter $\epsilon$ constrains 
a combination of the background magnetic field and the axion-photon coupling 
\cite{OstmMorts,MirRafSerp,ABRVJ} (and at the same time, through its relation (\ref{beta_eps}) 
to $\beta$ it places an indirect constraint on the variation of $T(z)$ as discussed 
above).  Intergalactic magnetic fields have not been directly observed but there are 
hints that they exist \cite{Kronberg:1993vk}.  For further discussion and relevant constraints, 
see \cite{Barrow:1997mj,Blasi:1999hu}.  Finally, note that axions can also convert back into 
photons -- a second order effect which could be significant for large mixing probability.  
 
Let us consider the constraint on $\beta$ coming from the distance duality 
bounds of Refs.~\cite{AVJ,ABRVJ}.  With $\beta=-2\epsilon/3$, combining the most recent
SN and $H(z)$ data (namely, the SCP Union2 Compilation \cite{Union2} together 
with the latest $H(z)$~\cite{SJVKS} data and Hubble parameter 
determination~\cite{Riess11}) yields 
\be\label{betaconstr}
\beta=0.01\pm0.04\,,
\ee
at 95\% confidence.

\begin{figure}
  \begin{center}
    \includegraphics[height=2.8in,width=3in]{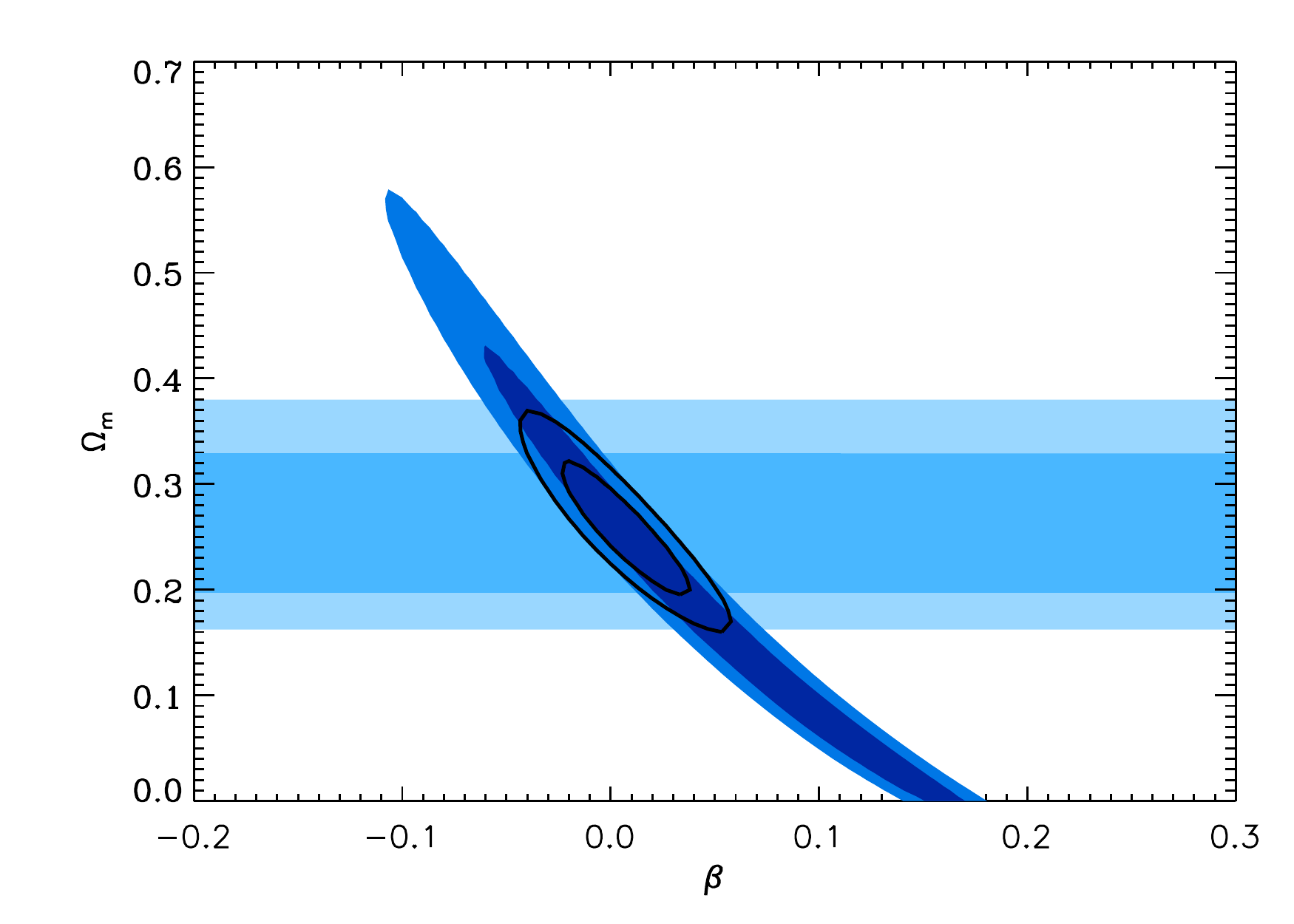}
    \includegraphics[height=2.8in,width=3in]{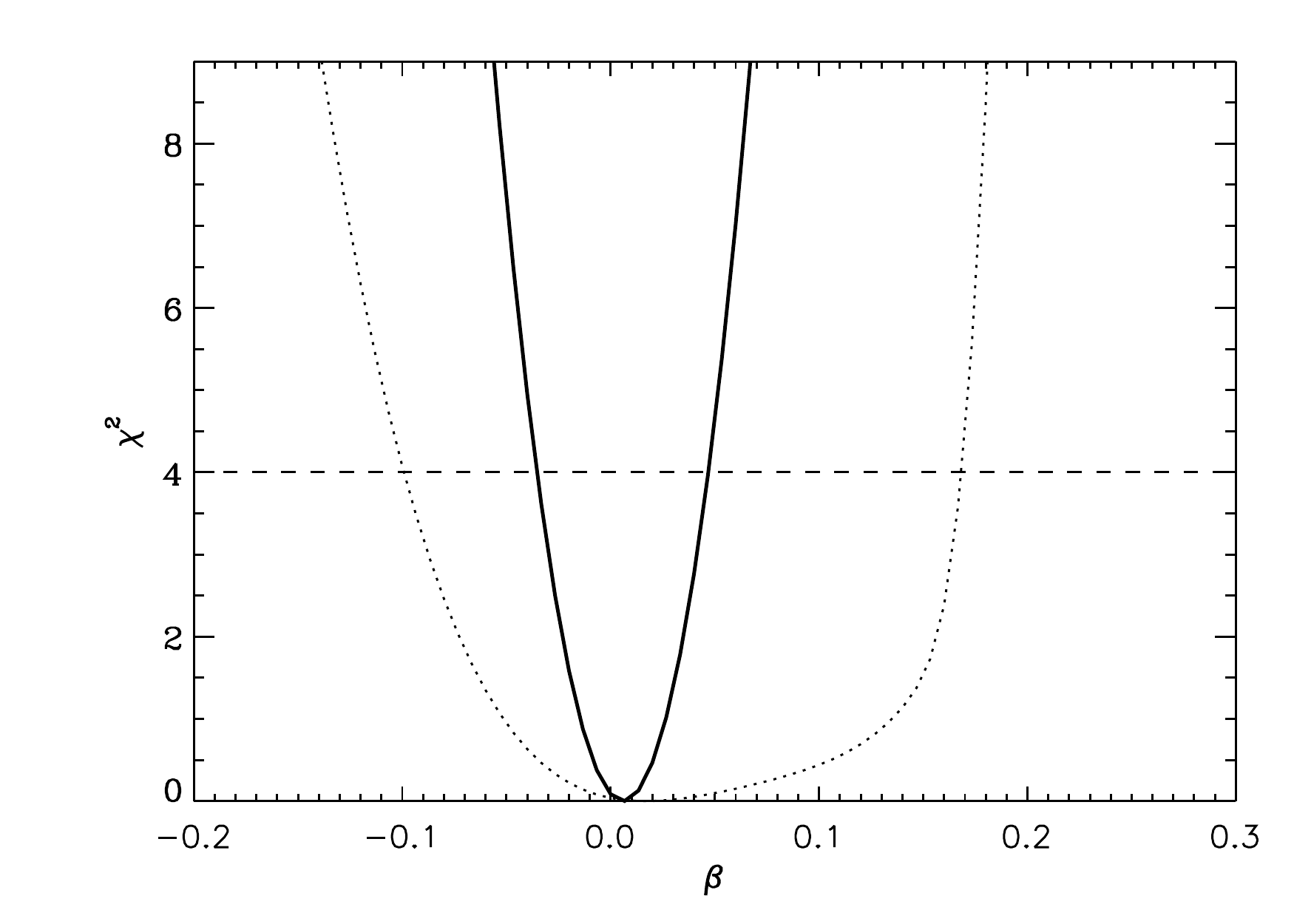} 
    \caption{\label{beta_constrs} 
                   Constraints from SN+$H(z)$ on the parameter 
                   $\beta$, parameterising violations of the temperature-redshift relation 
                   as $T(z)=T_0(1+z)^{1-\beta}$.  
                  {\it Left:} Two-parameter constraints on the $\beta-\Omega_m$ 
                  plane.  Dark blue contours correspond to 68\% and 95\% 
                  confidence levels obtained from SN data alone, light blue 
                  contours are for $H(z)$ data, and solid line transparent contours 
                  show the joint SN+$H(z)$ constraint.                   
                  {\it Right:} One-parameter joint constraints on $\beta$ marginalised 
                  over cosmological parameters. The solid line is for SN+$H(z)$ while 
                  the dotted one for SN data only.  The dashed line shows the $\Delta\chi^2=4$ 
                  level.}
  \end{center}
\end{figure}

This has been obtained by considering flat $\Lambda$CDM 
models and marginalising over the matter density $\Omega_m$ and Hubble 
parameter $H_0$.  Fig. \ref{beta_constrs} shows the relevant constraint on the 
$\beta-\Omega_m$ plane after marginalising over $H_0$ (left) and the constraint 
(\ref{betaconstr}) on $\beta$, obtained by marginalising over both $H_0$ and 
$\Omega_m$.  On the left, the dark blue contours are the $1$ and $2$-$\sigma$ 
joint (2-parameter) confidence levels for the SN data, the lighter blue regions show 
the corresponding constraints from $H(z)$ data, and the solid black lines show the 
combined SN+$H(z)$ constraints. Again, as we will discuss in \S\ref{forecastsdistance},
there are very good prospects for future improvements.

\section{\label{forecasts}Forecasts}

Having discussed the current constraints on the parameter $\beta$, we now move on to study in detail the prospects for improvements coming from the next generation of space and ground-based experiments. We will discuss in succession three different probes, which together span the redshift range $z\sim0-4$. By combining data from different observations one can therefore explore thoroughly a wide range of redshifts and significantly reduce the statistical uncertainties on the underlying phenomenological parameters. Moreover, given the very different nature of each type of measurements, one also has a much better control over possible systematics---this is particularly important in the event of a detection of deviations from the canonical behaviour.

\subsection{\label{forecastsPlanck}Low redshifts: Planck HFI (clusters)}

Planck HFI \cite{PLHFI2010Lamarre} was specifically designed from the beginning to measure the SZ effect in galaxy clusters \cite{Nabila1997}:               
the spectral coverage allows one to explore the positive and negative part of the spectral distortion,
and it is optimally suited for cluster detection and to break cluster parameters degeneracy.

The full sky survey will provide us with a SZ catalog of thousands of clusters. The Planck early results \cite{ESZ} already 
provide us with a sample of $189$ cluster candidates. 
Here we focus on a survey dedicated to a sample of well known clusters, for which X-ray and optical 
information is available, so dealing with a subsample of the final Planck cluster catalog (ESZ).                                                         

A catalog of 166 clusters has been built by using BAX (X-ray Clusters Database) \cite{BAX}:\\
\newline
\begin{tabular}{p{4.0cm}p{8cm}}
  \hline \hline
  \textbf{Cluster name} &  \\

  \textbf{RA (J2000)} & Right ascension\\

  \textbf{DEC (J2000)} & Declination\\

  \textbf{z} & redshift\\

  \textbf{$\rm F_{X}$} & unabsorbed X-ray flux in ROSAT band (0.1-2.4) keV in units of $(10^{-12} \rm erg/s/cm^{2})$
  \\

  \textbf{Reference-$\rm F_x$} & \\

  \textbf{$\rm L_{X}$} & X-ray luminosity in the ROSAT band (0.1-2.4) keV in $10^{44}$ ergs $s^{-1}$ \\

  \textbf{Reference-$\rm L_x$} & \\

  \textbf{Band-Inf} (keV)& \\

  \textbf{Band-Sup} (keV)& \\

  \textbf{$\rm T_{X}$} & X-ray gas temperature in keV\\

  \textbf{$\sigma_{\rm T_{X}}$} & \\

  \textbf{Reference-$\rm T_x$} & \\

  \textbf{Instrument} & \\

  \textbf{$\rm R_{core}$} & Core Radius (arcsec)\\

  \textbf{$\sigma_{\rm R_{core}}$} & \\

  \textbf{Reference-$R_{core}$} & \\
  
  \textbf{$\beta$} & slope of the gas density profile derived from the $\beta$-model fitting\\

  \textbf{$\sigma_{\beta}$} & \\ \hline
  \\
\end{tabular}
\\

To avoid confusion, note that this $\beta$-model does not refer to the same $\beta$ parameter as is discussed in the rest of the paper.

For each cluster we derive the following parameters:\\
\newline
\begin{tabular}{p{4.0cm}p{8cm}}
  \hline \hline \textbf{$n_{e0}$} &  central electronic density, assuming a isothermal $\beta$-model and following 
  \cite{furuzawa1998}\\

  \textbf{$y_{th}$} & central Comptonisation parameter \\

  \textbf{$\tau_{th}$} & central optical depth\\

  \textbf{$Y_{int}$} & Comptonisation parameter integrated over the cluster extent\\

  \textbf{$\rm D_{A_{SZX}}$} & Angular distance\\
  \hline
\end{tabular}
\\

We have simulated the observations of a sample of about 40 well known clusters, already observed by Planck \cite{ESZ}, 
taking into account the Planck HFI instrumental characteristics and observing strategy \cite{bluebook}. The frequency bands 
considered in the simulation are the four at lower frequencies (100, 143, 217, 353 GHz), since the
remaining two higher frequency bands (545, 857 GHz) are best suited for foregrounds extraction and are not 
useful for the reconstruction of the SZ signal. CMB and foregrounds are assumed as previously removed. 

The forecast for the SZ effect signal for the clusters has been obtained from the measured X-ray 
properties, assuming an isothermal model. The ICM pressure profile has historically been described by an isothermal $\beta$-model \cite{Cavaliere1978}. 
Recent X-ray observations have shown that a $\beta$-profile for gas density is a poor approximation for the cluster's profile at large radii, leading several authors to propose more realistic profiles \cite{Nagai2007, Arnaud2010}.
The use of the $\beta$-model in this work is for consistency with the X-ray derived parameters collected in the catalogue, 
which almost invariably were based on a $\beta$ profile assumption for gas.
Nevertheless, the analysis procedure to determine T(z) makes use of the central $\tau$ values, and it will not be affected by a different adopted profile.

We have used the Planck noise model (NET values as reported in \cite{bluebook}) 
to estimate the errors in the observed spectra. The integration time for 
the mock observation is obtained assuming uniform sky coverage and two years of observation. For each cluster 
the observation time is 10s. The dilution effect has been taken into account to estimate the error on the SZ signal for each channel.

The mock dataset was then analysed to recover the original input parameters of the cluster. The analysis 
has been performed through a Monte Carlo Markov Chain (MCMC) algorithm which allows us to explore the full space 
of the cluster parameters (optical depth $\tau$, peculiar velocity $v_{pec}$, electron temperature $T_e$) and 
the CMB brightness temperature at the redshift of the cluster. In the analysis we allowed
for calibration uncertainty, considered as an uncertain scale
factor, which was modelled as a Gaussian with mean 1 and $0.1\%$ standard deviation\footnote{The 
adopted value for the absolute calibration accuracy is in a way too optimistic with respect to HFI 
early maps ($\leq 2 \%$) \cite{PLHFIdataproc}, but still very conservative with respect to expected 
performances \cite{tristram2011}.  However, the impact of absolute calibration accuracy will mainly 
influence the $\tau$ parameter estimation, since $\Delta I$ depends linearly on $\tau$ at the first order.}.

The MCMC generates random sequences of parameters, which simulate posterior distributions for all parameters 
\cite{lewis2002}. The sampling
approach we used is the one proposed by Metropolis and Hastings
\cite{metropolis1953, hastings1970}. Convergence and mixing of the MCMC runs was 
tested through the Gelman-Rubin test \cite{gelman1992}.
We included a prior over the cluster gas temperature, as provided by X-ray data. 
For the radial component of the cluster peculiar velocities the prior is a theoretical one, a Gaussian with a 
universal vanishing mean and with a $1000$ km/s standard deviation. 
Clusters with almost flat $\tau$ posterior are excluded from the sample (this is the reason for which the sample is different
when the kinematic component is included from that when it is not included).

\begin{figure}
 \centering
 \includegraphics[width=10cm,keepaspectratio]{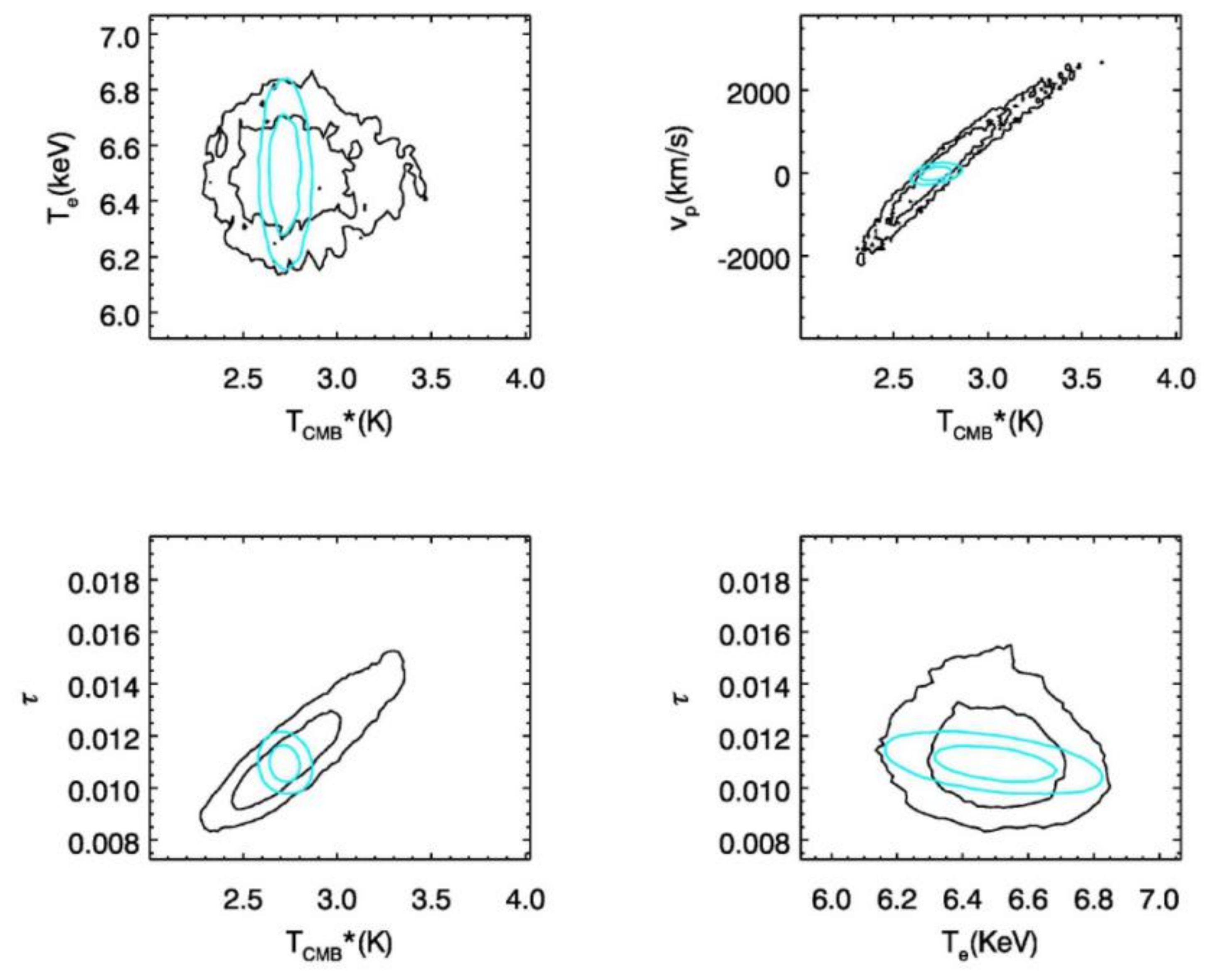}
 \caption{Parameter correlations for a simulated cluster: the
  contours show the $68\%$ and $95\%$ confidence limits from the
  marginalised distributions. Black contours are obtained allowing
  for a peculiar velocity prior with vanishing mean and standard
  deviation 1000 km/s; cyan contours are obtained allowing for a
  peculiar velocity prior with vanishing mean and standard deviation
  100 km/s. $T^{\ast}_{\rm CMB}$ corresponds to $T(z)/(1+z)$}   
\label{fig:contours}
\end{figure}

As was already noted in \cite{Luzzi} there is a degeneracy between $T_{\rm CMB}(z)$ and $v_{pec}$, see 
Fig. \ref{fig:contours}. In order to reduce the impact of this degeneracy and then to reduce the uncertainty in the determination of 
$T_{\rm CMB}(z)$, a better knowledge of the peculiar velocity is required or it is necessary to remove the 
kinematic component from the thermal component, together with the CMB intrinsic anisotropy. 
With only Planck measurements, given the multifrequency coverage, we can envisage the second option.
Kinematic SZ (KSZ) and the intrinsic CMB anisotropy have the same spectral shape in the non relativistic limit;
to separate the two components in nearby clusters or in distant clusters using beams larger than few arcmin
it is necessary to rely on the very small spectral distortions of the kinematic SZ due to relativistic effects \cite{Carlstrom2002}.
For Planck it is safe to consider that in a large majority of cases the spectral distortions of KSZ due to relativistic effects are 
too small with respect to CMB confusion and noise level to allow disentangling KSZ and CMB intrinsic anisotropy, i.e. the 
kinematic component is removed as a first step in cleaning maps of CMB contribution.

Nevertheless, in the following we will consider both cases, thus fitting either four parameters ($\tau$, $T_e$, $v_{pec}$, $T_{\rm CMB}(z)$) 
or three parameters ($\tau$, $T_e$, $T_{\rm CMB}(z)$).

\begin{figure}
 \centering
 \includegraphics[width=11cm,keepaspectratio]{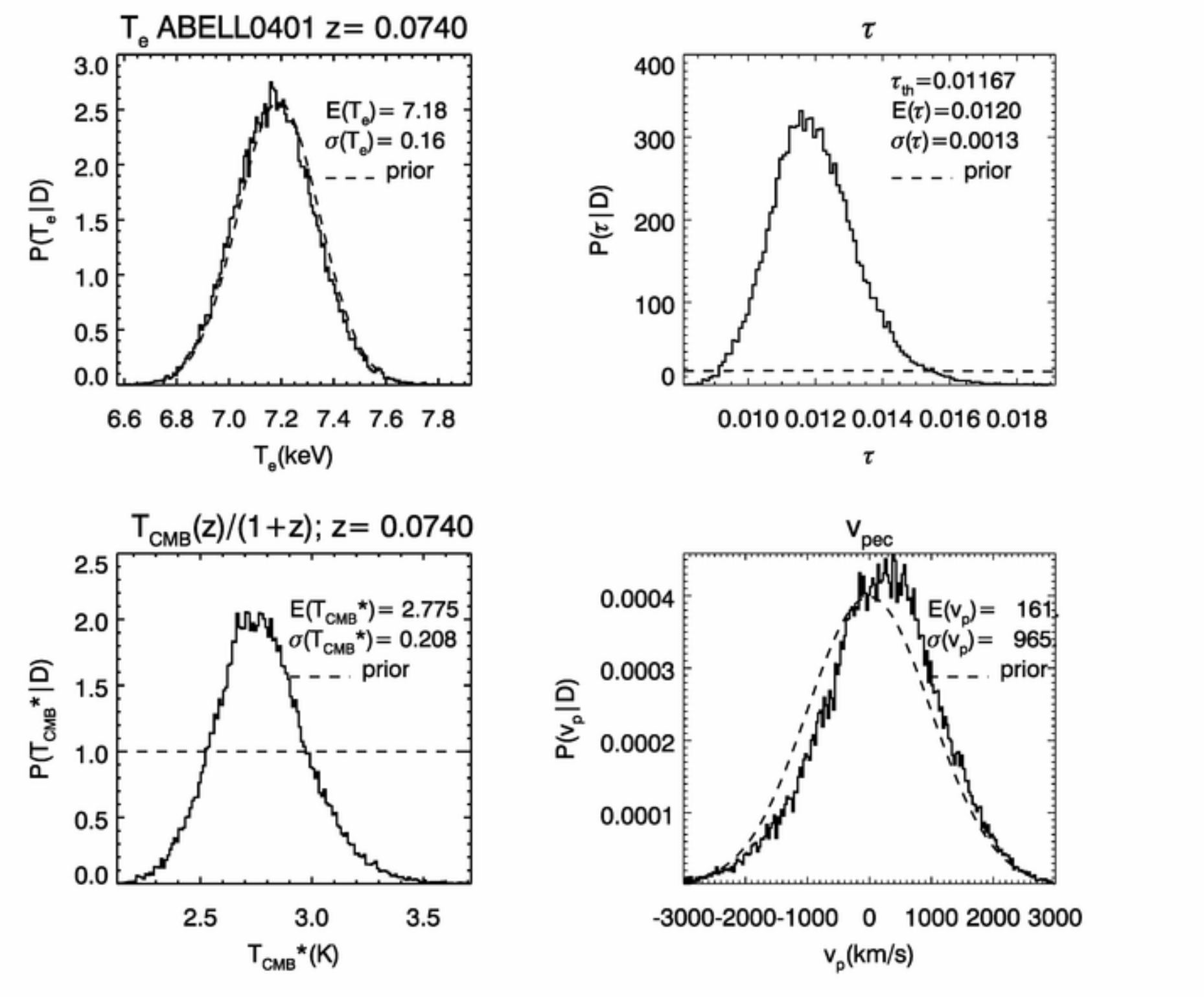}
 \caption{An example of parameter extraction for a single cluster, with kinematic component included.}
 \label{fig:singleclusterkin}
\end{figure}

\begin{figure}
 \centering
 \includegraphics[width=11cm,keepaspectratio]{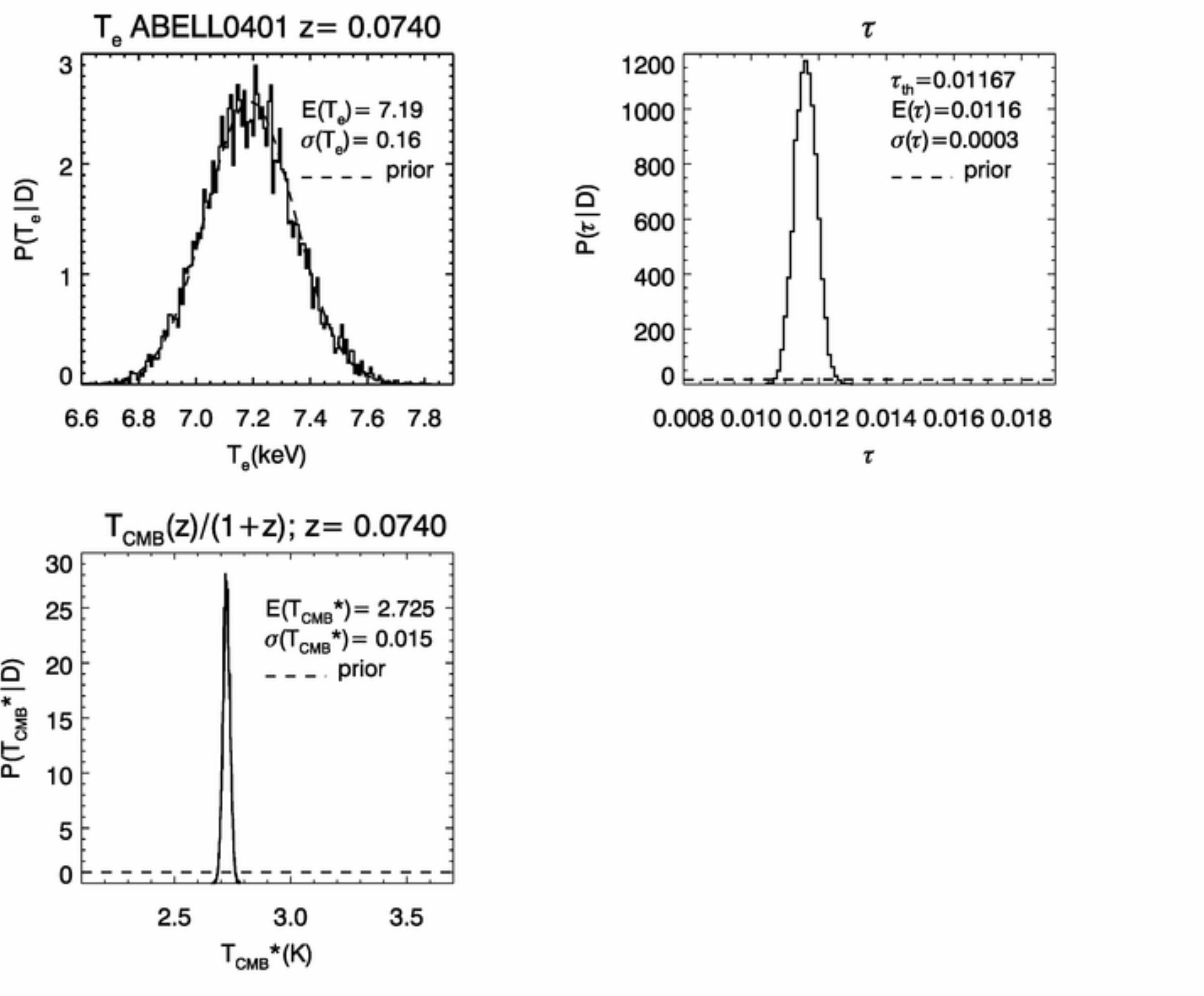}
 \caption{An example of parameter extraction for the same cluster of Fig. \ref{fig:singleclusterkin}, with kinematic component not included.}
 \label{fig:singleclusternokin}
\end{figure}

\begin{figure}
 \centering
 \includegraphics[width=10cm,keepaspectratio]{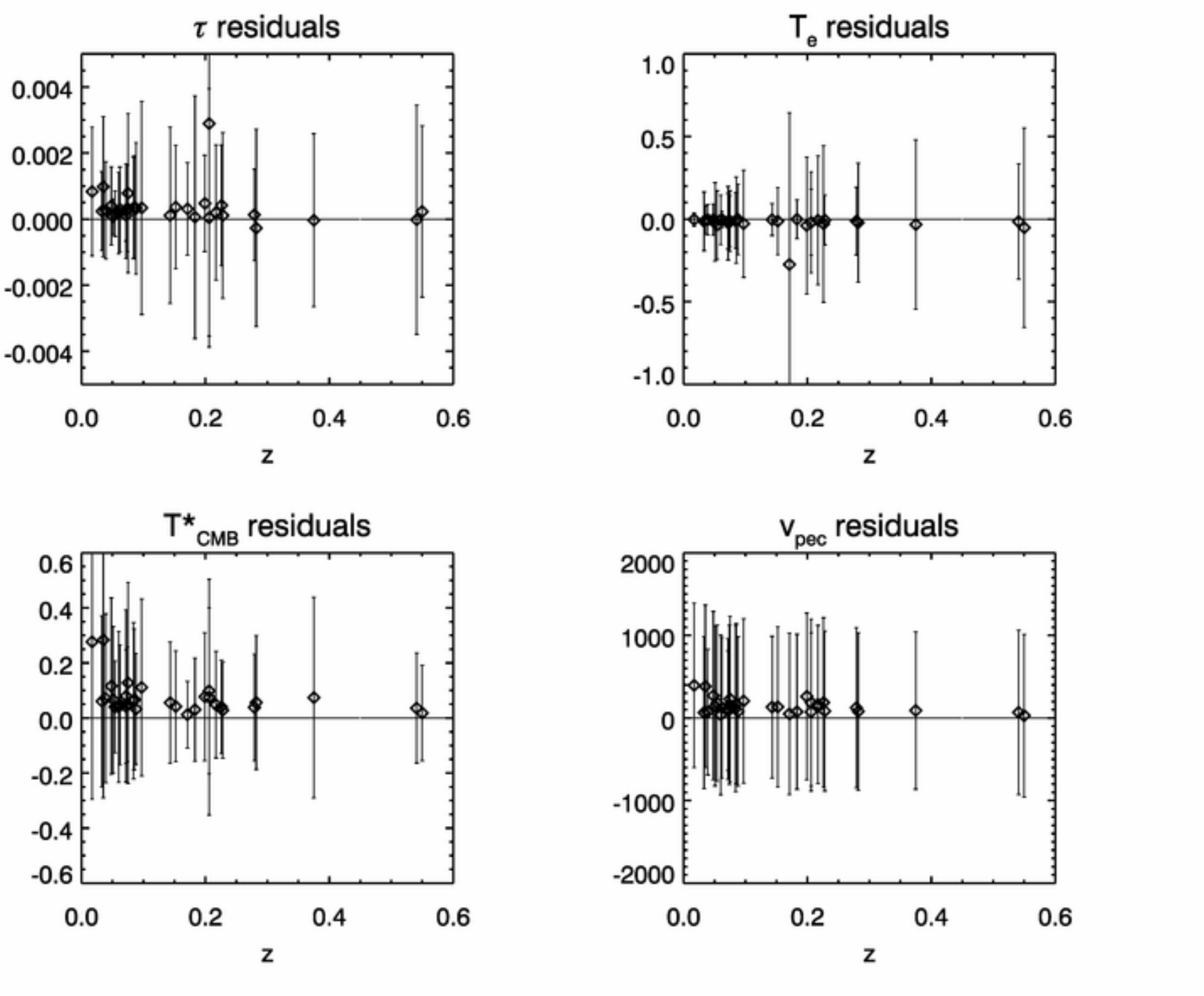}
 \caption{Residuals for $32$ clusters, with kinematic component included. All the clusters with almost flat $\tau$ posterior have been excluded from the original $42$-cluster sample.}
 \label{fig:residualswithksz}
\end{figure}

\begin{figure}
 \centering
 \includegraphics[width=10cm,keepaspectratio]{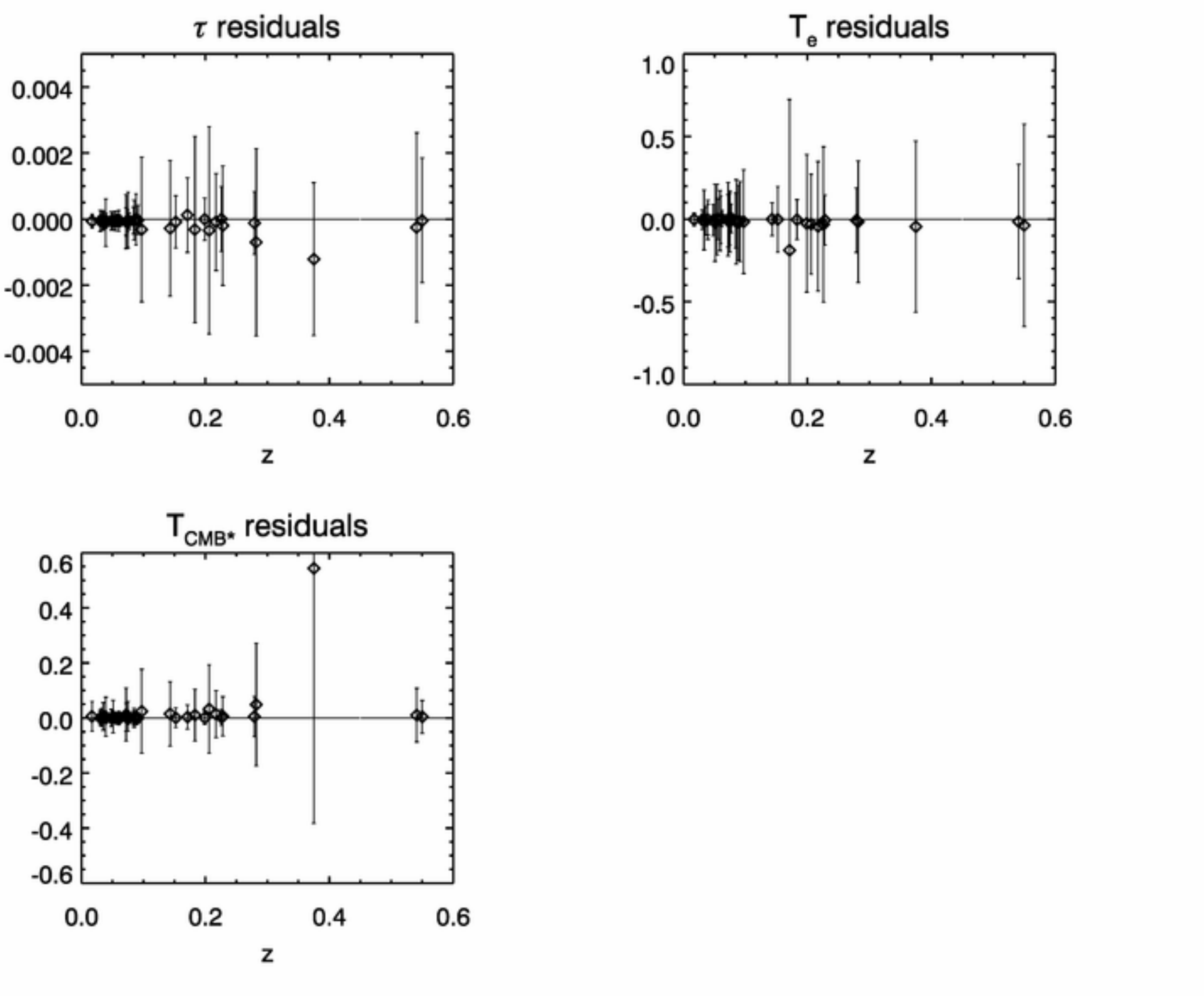}
 \caption{Residuals for $37$ clusters, with kinematic component previously removed. All the clusters with almost flat $\tau$ posterior are excluded from the original $42$-cluster sample.}
 \label{fig:residualswithnoksz}
\end{figure}

In Figs. \ref{fig:singleclusterkin} and \ref{fig:singleclusternokin} we present the results of the parameter estimation
analysis for a single cluster, both with and without the kinematic component. 
In Figs. \ref{fig:residualswithksz} and \ref{fig:residualswithnoksz} we show the residuals for cluster parameters 
for the whole sample.

\newpage

%\newpage

\begin{figure}
 \centering
 \includegraphics[width=7cm,keepaspectratio]{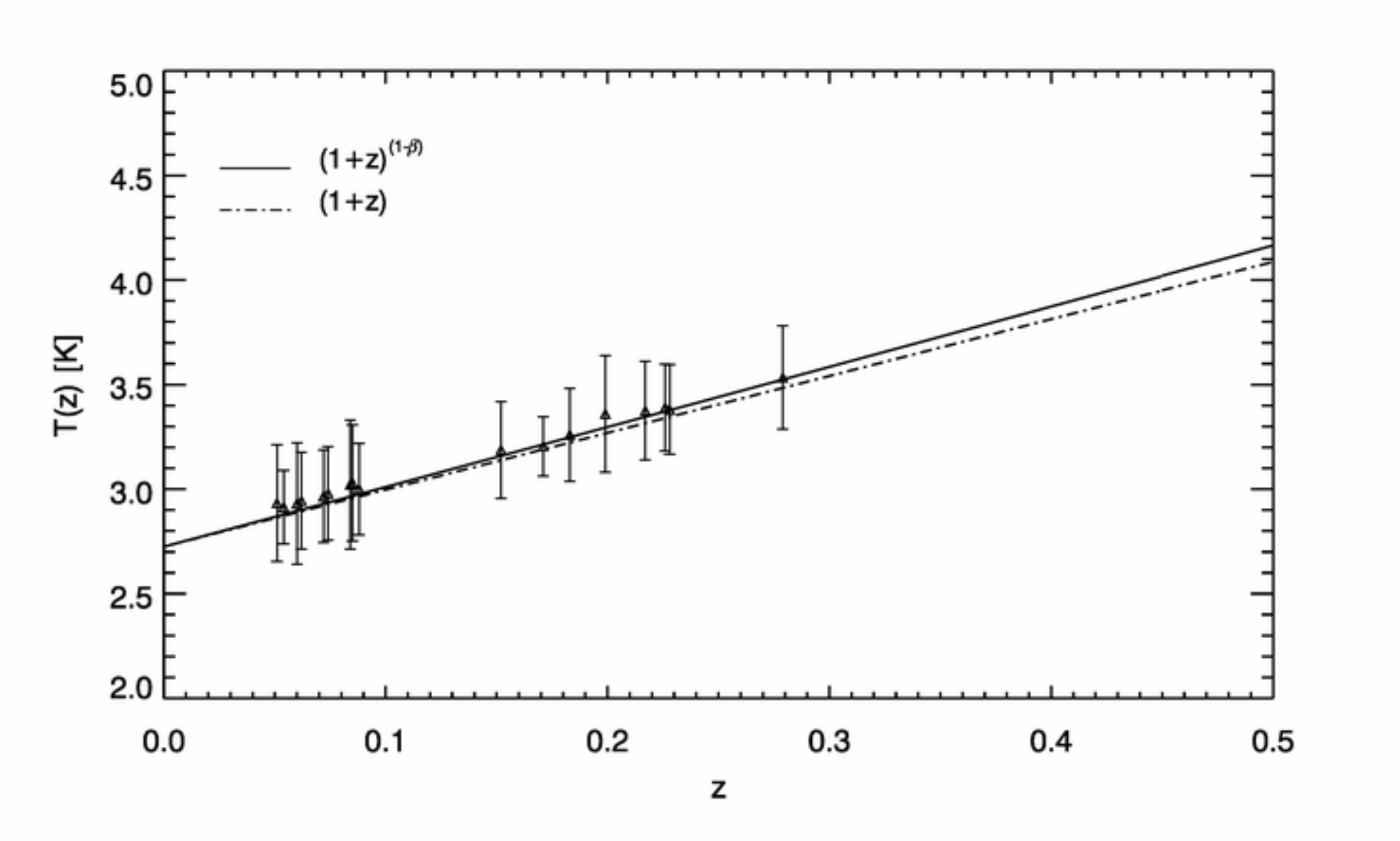}
 \includegraphics[width=4cm,keepaspectratio]{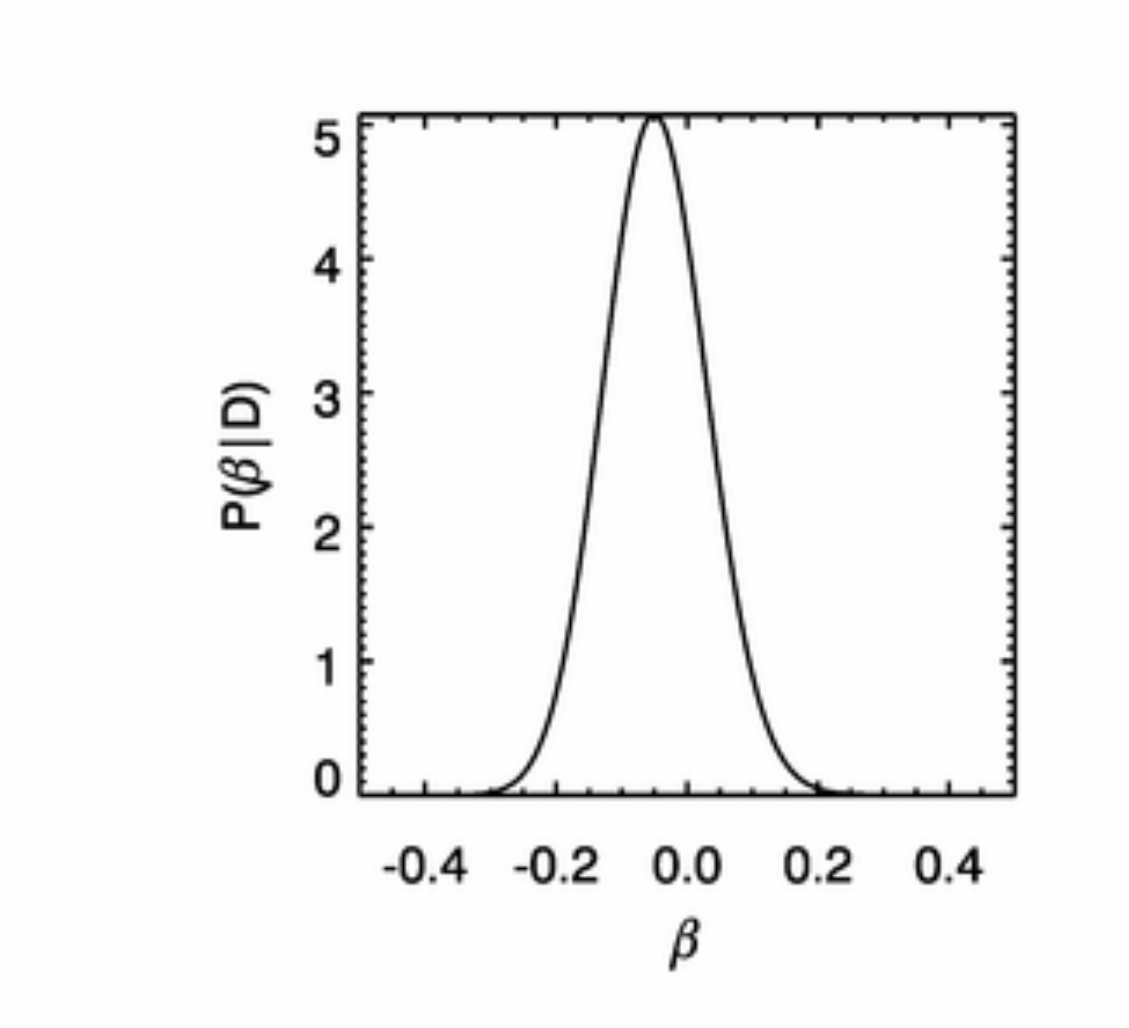} 
 \caption{Left: $T_{\rm CMB}$ vs z, with the kinematic component included. Right: Posterior of the $\beta$ parameter, as obtained by performing a fit of the $T(z)$ data points.}
 \label{fig:tcmbvszkin}
\end{figure}

\begin{figure}
 \centering
 \includegraphics[width=7cm,keepaspectratio]{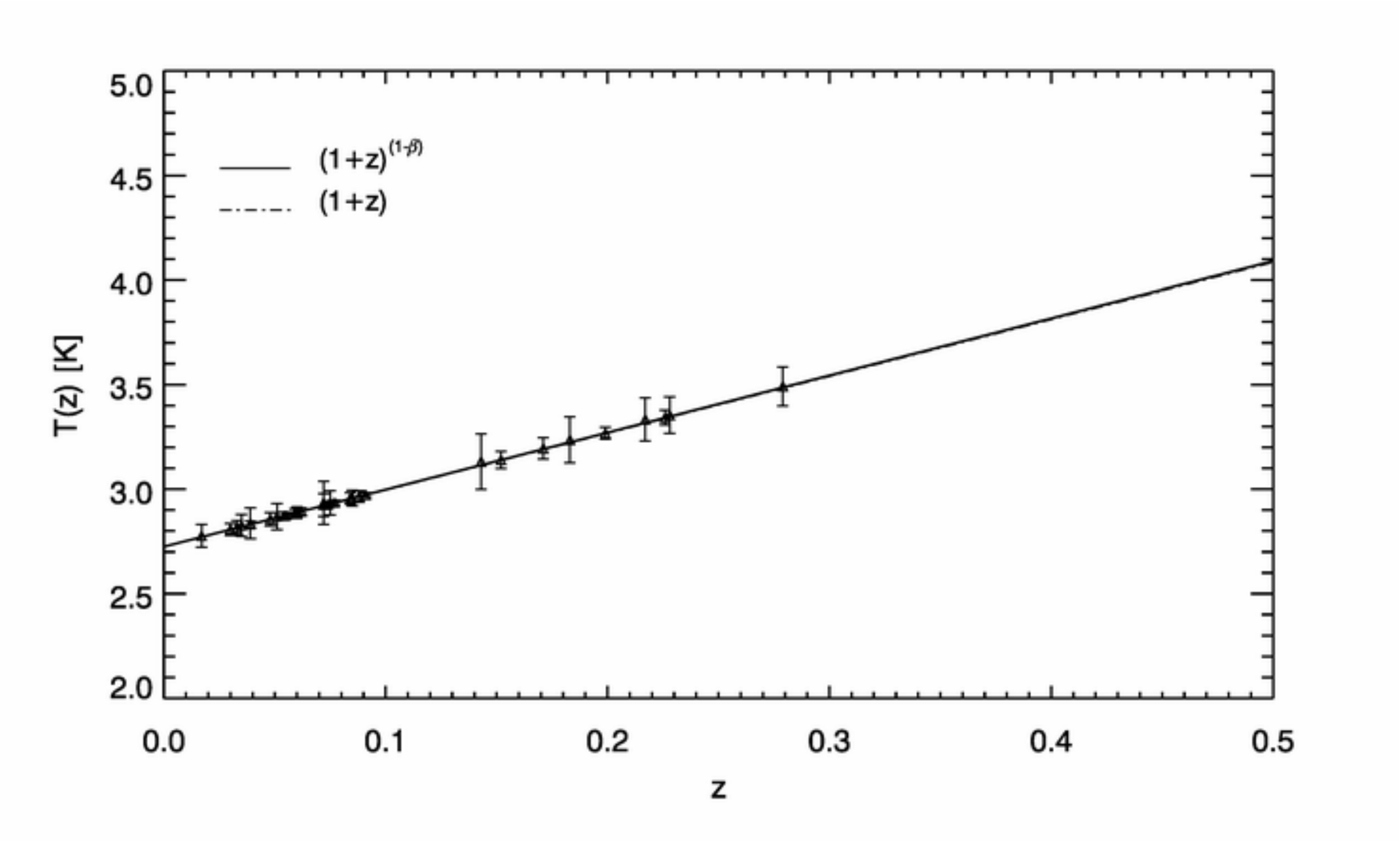}
 \includegraphics[width=4cm,keepaspectratio]{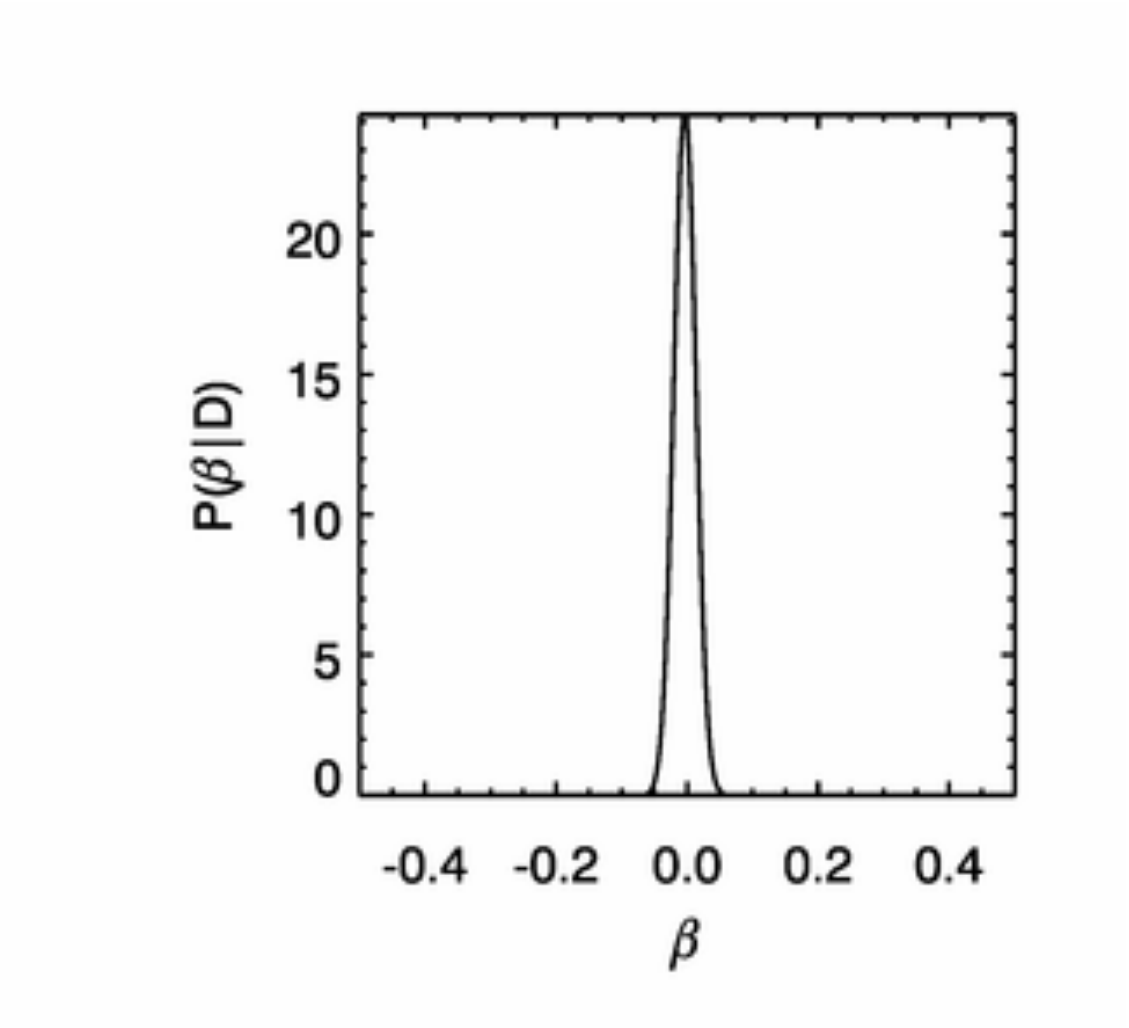} 
 \caption{Left: $T_{\rm CMB}$ vs z, with the kinematic component not included. Right: Posterior of the $\beta$ parameter, as obtained by performing a fit of the $T(z)$ data points.}
 \label{fig:tcmbvsznokin}
\end{figure}

To obtain $\beta$ we have performed a fit of the $T(z)$ data points (see Figs. \ref{fig:tcmbvszkin} and
\ref{fig:tcmbvsznokin}). The final $\beta$ value we get by fitting the $T(z)$ data points obtained
with the MCMC treatment is $\beta = -0.047 \pm 0.079$ when the kinematic component is included (from a
final sample of 19 clusters, selected with the condition of non flat $\tau$ posterior and $S/N \geq 6$)
and $\beta = -0.003 \pm 0.016$ when the kinematic component is previously removed (from a final sample of $37$
clusters, selected with the condition of non flat $\tau$ posterior and $S/N \geq 6$). 

\newpage

In conclusion, if the kinematic component is removed altogether with the CMB primary anisotropy component,  
with Planck we can reach $0.6\%$ sensitivity on $T_{\rm CMB}(z)$ measurements; otherwise the sensitivity
will be around $7\%$. With only tens of clusters we can in principle get better constraints on $\beta$
than the current results with SZ+Atomic carbon+CO.

\subsection{\label{forecastsdistance}Intermediate redshifts: EUCLID/SNAP (distance measurements)} 

In this section, we show forecast constraints on the temperature-redshift relation 
(in particular on the parameter $\beta$ of \S\ref{theory}) which could be achieved 
by combining $H(z)$ measurements from upcoming spectroscopic BAO surveys with 
future SN data. The next decade will see a dramatic improvement on $H(z)$ and 
angular diameter distance data at redshifts $z\lesssim 2$, notably through ongoing 
and upcoming BAO surveys like BOSS \cite{BOSS} and EUCLID \cite{EUCLID}.  
Similarly, future SN missions (e.g. SNAP \cite{SNAP}) will dramatically reduce 
the errors in SN brightness data. 

As discussed in \cite{ABRVJ}, BOSS will not significantly improve opacity bounds (on 
which our constraint on $\beta$ are based) with respect to current $H(z)$ cosmic chronometer data (i.e.
from differential ageing of luminous red galaxies \cite{SJVKS}), because it will be restricted to redshifts $z\le 0.7$. 

On the other hand EUCLID -- a combination of the earlier SPACE \cite{SPACE} and DUNE \cite{DUNE} missions -- 
will reach much higher redshifts and is expected to dramatically improve these constraints. 
Aiming for launch in 2019, it would cover about 20,000 ${\rm deg}^2$ of sky providing 
around 150 million redshifts in the range $z<2$.  Here, we consider forecast constraints 
from EUCLID and a Supernova SNAP-like survey (or dark energy task force stage IV SNe 
mission) \cite{DETF}. 

\begin{figure}
  \begin{center}
    \includegraphics[height=3in,width=3in]{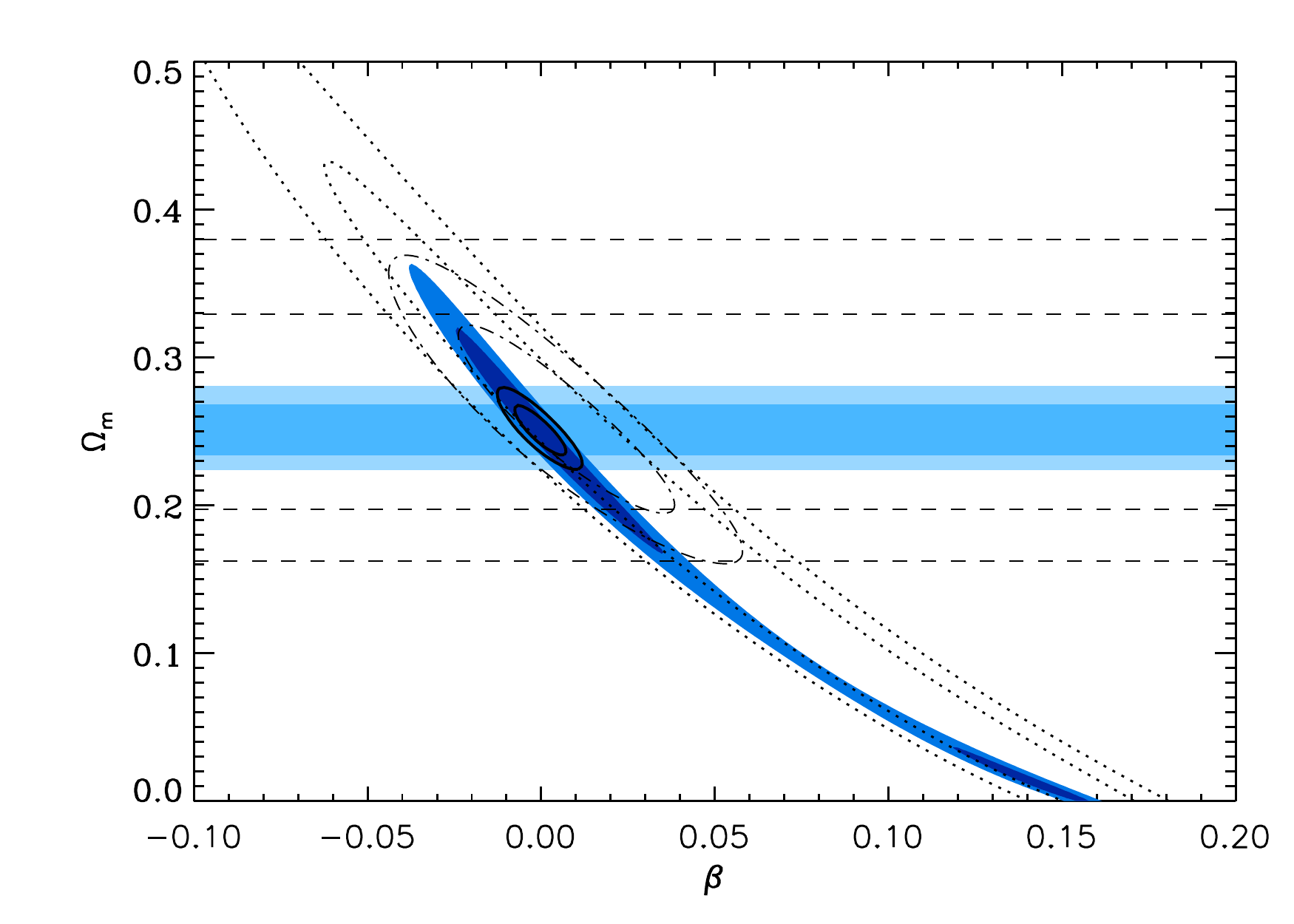} 
    \includegraphics[height=3in,width=3in]{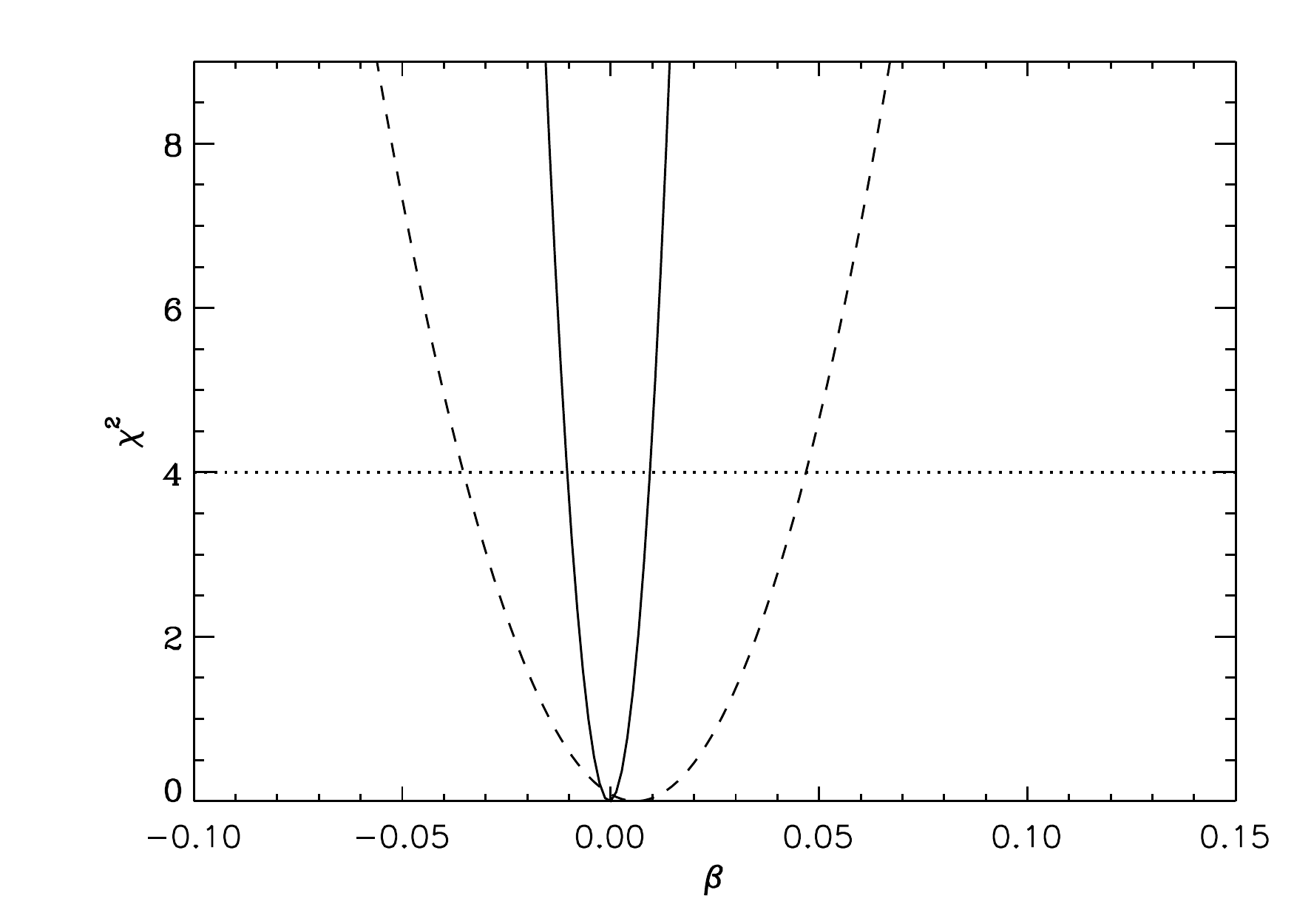} 
    \caption{\label{forecast_beta} Future constraints on the parameter 
                  $\beta$ from EUCLID and SNAP. 
                  {\it Left:} Two-parameter constraints on the $\beta-\Omega_m$ plane.  
                   Dark blue contours 
                   correspond to 68\% and 95\% confidence levels from SNAP alone, 
                   light blue contours are for EUCLID, and solid line transparent contours 
                   show the joint SNAP+EUCLID forecast constraint. Also shown are 
                   current constraints from $H(z)$ 'chronometer' data (dashed), 
                   SN data (dotted), and joint $H(z)$+SN (dot-dashed), presented 
                   in \S\ref{distance}, Fig.~\ref{beta_constrs}.                    
                  {\it Right:} One-parameter joint constraints on $\beta$ marginalised 
                  over cosmological parameters. The solid line shows the forecast 
                  constraint from EUCLID+SNAP, while the dashed line corresponds 
                  to the current constraint from $H(z)$+SN, discussed in 
                  \S\ref{distance}, Fig.~\ref{beta_constrs}. The dashed line is the 95\% 
                  confidence level.           
                  }
  \end{center}
\end{figure}

We use the code developed by Seo \& Eisenstein \cite{SeoEisen} to estimate
the errors in radial distances achievable by using BAO as a standard 
ruler.  Fig.~\ref{forecast_beta} shows our forecasted constraints on the parameter 
$\beta$, using modelled BAO data with forecasted errors for EUCLID, combined 
with modelled SN data and errors for a SNAP-like survey.

On the left panel, light
blue contours show the 1-$\sigma$ and 2-$\sigma$ (2-parameter) constraints on 
the $\beta$--$\Omega_m$ plane from EUCLID only, darker blue contours show the 
corresponding constraints from SNAP, while solid line transparent contours show
the (2-parameter) joint EUCLID+SNAP forecast constraints.  To make a more direct 
comparison we have also shown the corresponding constraints obtained from 
current data, namely `cosmic chronometer' $H(z)$ (dashed), SN (dotted), and 
joint $H(z)$+SN (dot-dashed), discussed in \S\ref{distance}.  The right panel 
shows the relevant 1-parameter constraint on $\beta$, after marginalising over 
$\Omega_m$, for EUCLID+SNAP (solid) and current $H(z)$+SN data (dashed).  
The dotted line is the 95\% confidence level, $\Delta\chi^2=4$.   

Overall, the improvement with respect to current constraints by combining 
EUCLID+SNAP will be quite significant, with the area of the joint constraint 
in Fig.~\ref{forecast_beta} (left) reduced by a factor of a few decades.  The 
one parameter constraint on $\beta$ will be improved by a factor $\gtrsim 5$,
reaching $|\Delta\beta|\sim 0.008$ at 95\% confidence. This is competitive 
to the Planck HFI result, discussed above.

\begin{figure}
 \centering
 \includegraphics[width=11cm,keepaspectratio]{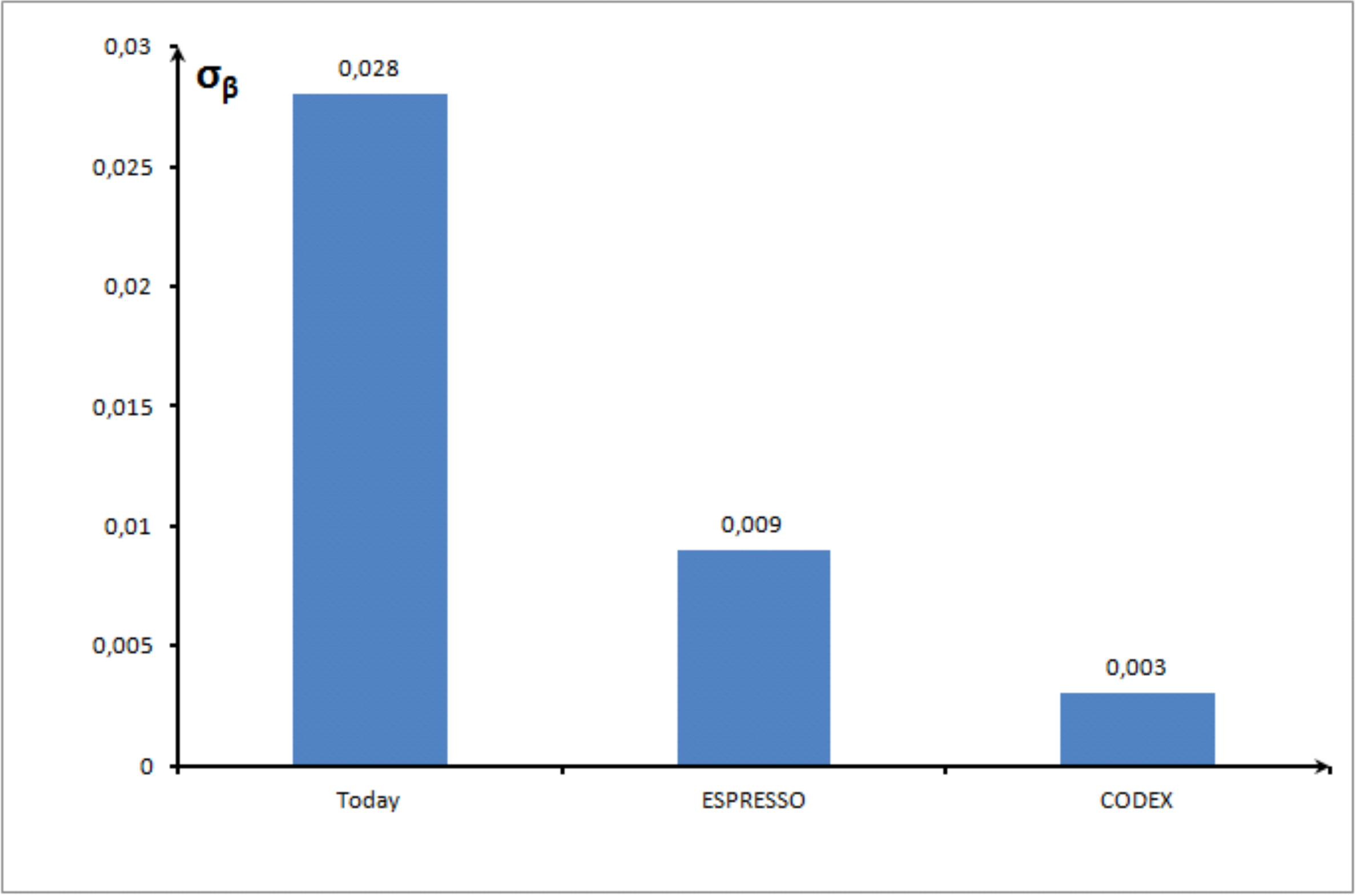}
 \caption{Forecasts for direct constraints on the phenomenological parameter $\beta$ for ESPRESSO and CODEX, compared with the current uncertainty (coming from combining all available direct measurements, cf. \S\ref{SZ}).}
 \label{fig:codex}
\end{figure}

\subsection{\label{forecastspectroscopy}High redshifts: ESPRESSO and CODEX (spectroscopy)} 

ESPRESSO\footnote{See http://espresso.astro.up.pt/} (for the VLT) and CODEX\footnote{See http://www.iac.es/proyecto/codex/}
(for the E-ELT) are two forthcoming ESO high-resolution, ultra-stable spectrographs. Although
their common cosmology-related science driver is the precise spectroscopic measurement of nature's
fundamental couplings (particularly the fine-structure constant $\alpha$ and the proton-to-electron
mass ratio $\mu$, see \cite{JENAM}), they will be in a unique position to carry out precise
measurements of $T(z)$ at high redshift.

As discussed in \S\ref{SPC}, there are currently 5 CO absorption systems, in the redshift range $z\sim1.5-3.0$, where the CMB temperature can be measured with an uncertainty
\begin{equation}
\Delta T_{\rm Now}\sim0.7\, K\,.
\end{equation}
Based on current plans for ESPRESSO and CODEX \cite{Cristiani,Liske,SPIE}, we can estimate that they will be able to reduce this uncertainty to, respectively,
\begin{equation}
\Delta T_{ESP}\sim0.35\, K
\end{equation}
and
\begin{equation}
\Delta T_{COD}\sim0.07\, K\,.
\end{equation}
Given the planned redshift range of both spectrographs, their measurements will on average be done at higher redshifts, and here we will assume a redshift range $z\sim2.8-4.0$; from a theoretical point of view, going to higher redshifts is obviously desirable since they provide a bigger lever arm; however, one also has to keep in mind that these systems will be fainter.

Having said that, it is important to realise that the bottleneck here is not the amount of telescope time required to observe these systems (although that naturally grows as systems become fainter). In fact, in some systems that are observed with the aim of measuring $\mu$ one can also measure $T(z)$, so there could in fact be no extra cost in terms of telescope time. Instead, the bottleneck is simply finding more systems where these measurements can be made. Ongoing surveys such as SDSS-III BOSS \cite{BOSS} can play an important role in this endeavour. For the purposes of the present analysis we will assume that ESPRESSO will accurately measure $T(z)$ in 10 systems, while CODEX will measure 20 systems.

One can then generate mock catalogs of $T(z)$ measurements, assuming the standard scenario as a fiducial model and the above temperature uncertainties and redshift ranges, and determine the constraint on $\beta$ that such a catalog can yield. For simplicity we also assume a uniform probability in the redshift distribution of the sources. We will also assume that the uncertainty in the measurement at $z=0$ remains unchanged. Naturally, the constraint will have a mild dependence on where in the allowed redshift range the small number of sources happen to fall, but by generating a large number of realisations one can infer a representative uncertainty on $\beta$.

The results of this analysis are summarised in Fig.~\ref{fig:codex}: ESPRESSO is expected to improve on the constraints coming from currently available spectroscopic measurements by a factor of 3, and CODEX should improve on ESPRESSO by another factor of 3. As stated above, if more systems are found where the spectrographs can make these measurements, one could in principle further improve the sensitivity on $\beta$.

\section{\label{results}Summary: current and future constraints}

\begin{figure}
 \centering
 \includegraphics[width=11cm,keepaspectratio]{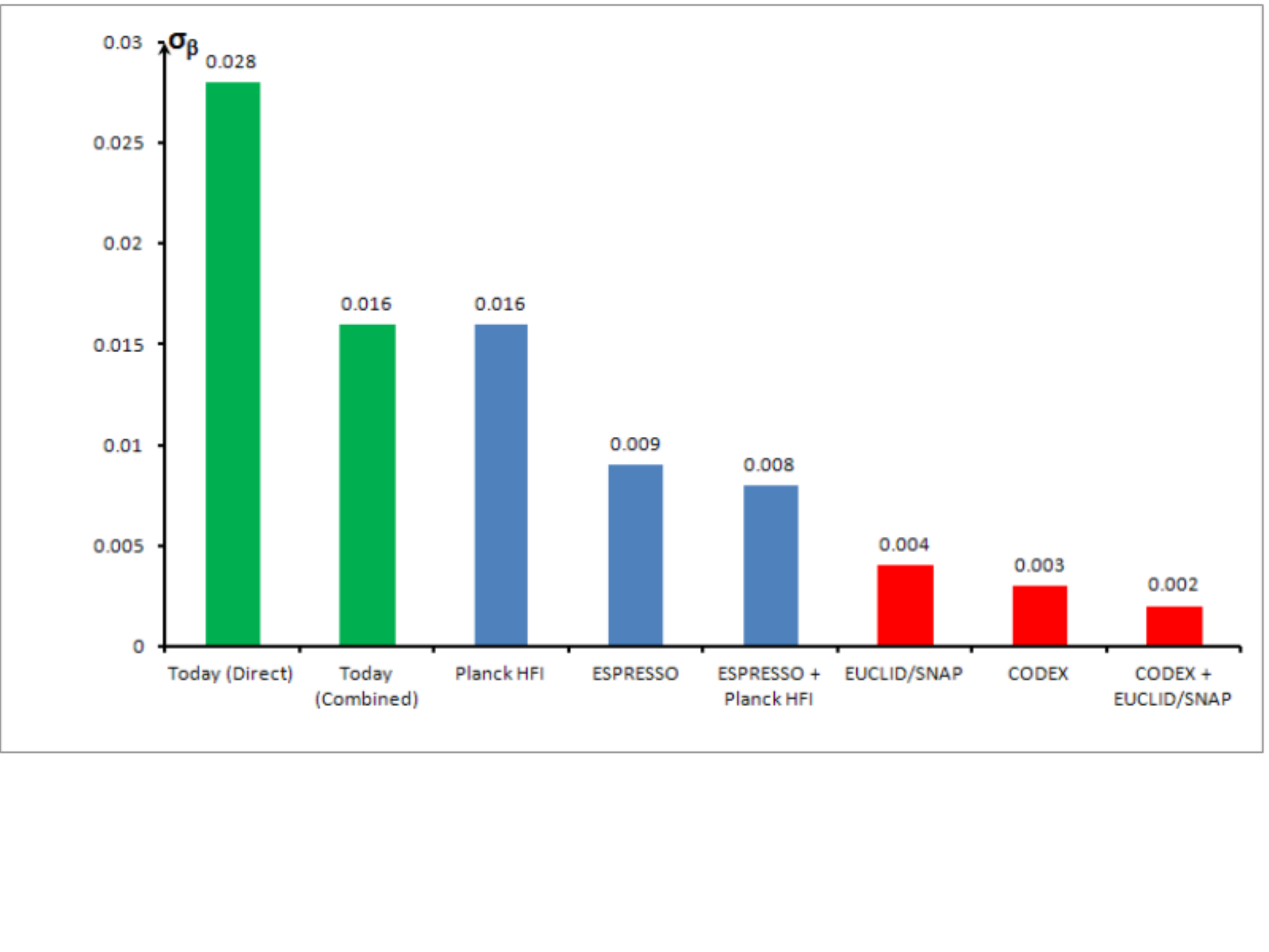}
 \caption{Comparing the current one-sigma uncertainties on the parameter $\beta$ (from direct measurements and from a combination of direct and indirect ones) with that achievable by ongoing and future experiments, specifically Planck HFI and ESPRESSO in the near future and EUCLID/SNAP and CODEX in the longer term.}
 \label{fig:allsigmas}
\end{figure}

We are now in a position to summarise the constraints on allowed deviations from the standard temperature-redshift relation. By combining the direct constraints from Noterdaeme {\it et al.} \cite{Noterdaeme} with the indirect ones that we have obtained in \S\ref{distance}, we finally obtain the weighted mean result
\begin{equation}\label{combined_constr}
\beta=0.004\pm0.016\,,
\end{equation}
which is a $40\%$ improvement on the direct constraint. We note that the three observational methods (clusters, distance measurements and spectroscopy) are nicely complementary, not only in terms of possible systematic uncertainties but also in terms of the redshift ranges covered. These results are summarised in Fig.~\ref{fig:allsigmas}, and compared with the forecasts for the various future experiments that we discussed in the previous section.

For experiments whose results will be available before 2020, we have considered Planck HFI and the ESPRESSO spectrograph.
One can see that Planck alone should be able to do as well as all the current data, with only tens of clusters. When ESPRESSO becomes available, it will allow a gain of a factor of 3 in sensitivity relative to the current direct constraints, and a factor of 2 relative to the constraints available just before it.

We point out that the Early SZ Planck catalogue \cite{Douspis11} consists of $189$ clusters and it is obtained with a 
selection criterion of S/N$>$6, thus implying that the SZ cluster sample for which $T_{\rm CMB}$ can be extracted 
is at least a factor of 4 larger than the one presented in this work; assuming that the spanned redshift range is 
the same as the one taken into account here (thus sensitivity on $T_{\rm CMB}(z)$ measurements is almost unchanged) then we have a $50\%$ improvement on the constrained $\beta$ ($\sigma_{\beta}\sim 0.008$). 
In principle, spatially resolved spectroscopic observations of galaxy clusters (as proposed with SAGACE \cite{deBernardis2010} or Millimetron\footnote{See http:/www.sron.rug.nl/millimetron},  would allow to further improve these constraints.
Finally, we note that there are also other techniques that can in principle be used to obtain these constraints from SZ clusters \cite{Ivan}.

In the longer (post-2020) term, CODEX and a combination of EUCLID and (some form of) SNAP will bring significant further improvements. Here the order in which they will become available is uncertain (as are, to some extent, their detailed characteristics), but their combined results are expected to bring a gain of about an order of magnitude relative to current sensitivities and of factors between 2 and 4 relative to the constraints available just before them.

Our analysis may in some ways be too simplistic, but we emphasise that in other ways it is fairly conservative. This is particularly the case when it comes to the assumptions on the number of systems in which the measurements can be made. The Early SZ Planck catalogue is evidence of this fact for the low redshift range, but the same is true for the spectroscopic measurements at high redshift. Given the  exquisite resolution and stability of the forthcoming ESO spectrographs and the large redshift lever arm they can probe, the most cost-effective way to further improve these constraints is undoubtedly to identify further systems where ESPRESSO and CODEX can make these measurements.

\section{\label{concls}Conclusions}

In this paper we have explored novel techniques for constraining physics beyond 
the standard model, focusing, in particular, on cosmological scenarios that can 
violate photon number conservation.  These include -- but are not limited to -- models 
in which photons mix with axion-like particles, decaying vacuum cosmologies and 
other photon injection mechanisms, models with astrophysical dust, and so on.      
By noting that such models can simultaneously modify the cosmological distance 
duality relation and the CMB temperature scaling law (these modifications being 
parametrically related to each other), we have initiated a programme of studying 
the consistency of these models through a combination of direct and indirect probes.  
Relevant probes include SN brightness measurements (yielding luminosity distances), 
galaxy ageing and/or BAO techniques (giving radial and angular diameter distances), 
SZ measurements of galaxy clusters (providing $T(z)$ at low redshifts $0\!<\!z\!<\!1$)  
and quasar absorption line spectroscopy (measuring $T(z)$ at higher redshifts 
up to $z\sim$ a few).  
       
This significantly enlarges ones' toolbox for studying cosmological models beyond 
the standard paradigm and leads to a notable improvement on current constraints 
on such models.  Indeed, the probes used are complementary, each having different 
systematics and/or redshift cover, so combining data from several  
probes allows one to reduce systematic uncertainties and obtain more stringent 
consistency checks, as well as improved constraints on the models under study.  
The combined bound (\ref{combined_constr}), which we have obtained on the 
parameter $\beta$ quantifying deviations from the standard $T(z)$ law (see 
equation (\ref{Tofz})), is a $40\%$ improvement over the corresponding direct 
constraints in \cite{Noterdaeme}.                 

The potential of this programme is enormous.  Considering ongoing and future 
missions (Planck HFI, EUCLID, SNAP, ESPRESSO, CODEX) we have obtained 
forecast constraints/errors on the parameter $\beta$. 
The expected improvement 
in current 1-$\sigma$ error bars can be up to two orders of magnitude, as summarised 
in Fig.~\ref{fig:allsigmas}.  There is also room for expanding further the current toolbox 
by accommodating more probes, again with different systematics and redshift cover.  
For example, the position of the first acoustic peak in the CMB is also sensitive to
distance duality violations and so provides a new tool at the redshift of last 
scattering~\cite{inprogress1}.      

In this work, we have only tried to demonstrate the effectiveness of these techniques by 
considering simple parameterisations, which are not necessarily physically motivated.   
These may be adequate for work with currently available data, but as redshift cover 
increases and sensitivity improves, better parameterisations will be required.  
However, it is important to highlight that, even with current sensitivity, the techniques 
described here have a notable potential for constraining specific cosmological 
and high-energy physics models beyond the standard paradigm if specific `model-tailored'
parameterisations are used.  Within a given model, the underlying physics often points 
to specific parameterisations for the violation of standard laws, and these parameters 
can be directly related to fundamental/microphysical parameters of the underlying theory.  
This approach was initiated in reference~\cite{ABRVJ}, where specific parameterisations 
for distance duality violation were adopted for different models (photon-axion 
mixing, hidden photons, mini-charged particles).  Other scenarios that can 
be probed with the tools described in this paper, through the use of more 
realistic parameterisations as suggested by the theory, include varying 
$\alpha$ cosmologies and dynamical dark energy.  These will be discussed 
in detail in a follow-up publication~\cite{inprogress2}.

\section*{Acknowledgements}

This work was done in the context of the cooperation grant `Evolution and Astrophysical Consequences of Cosmic Strings and Superstrings' (ref. B-13/10), funded by CRUP and The British Council. We also acknowledge the support of Funda\c{c}\~ao para a Ci\^encia e a Tecnologia (FCT), Portugal, in the form of the grant PTDC/FIS/111725/2009.

A.A. was supported by a CTC Postdoctoral Fellowship at DAMTP, University of Cambridge 
and in part by a Marie Curie IEF Fellowship at the the University of Nottingham. 
The work of G.L. is funded by a CNRS Postdoctoral Fellowship at LAL, Centre Scientifique d'Orsay. 
The work of C.M. is funded 
by a Ci\^encia2007 Research Contract, supported by FCT/MCTES (Portugal) and POPH/FSE (EC).
The work of A.M.M. was partially funded by Grant No. CAUP-03/2011-BII.
  
This research has made use of the X-Rays Clusters Database (BAX) which is operated by the Laboratoire d'Astrophysique de Tarbes-Toulouse (LATT), under contract with the Centre National d'Etudes Spatiales (CNES). We also acknowledge useful discussions on $T(z)$ measurements with Paolo Molaro and Patrick Petitjean, particularly for defining the observational scenarios for ESPRESSO and CODEX.

\end{document}